\documentclass[twocolumn,english,aps,pra,10pt,superscriptaddress,floatfix]{revtex4-2}
\usepackage{times}
\usepackage{graphicx}
\usepackage{amssymb}
\usepackage{bm}
\usepackage{amssymb,amsfonts,amsmath,amsbsy,bm,t1enc,latexsym}
\usepackage{float}
\usepackage[colorlinks=true,citecolor=blue,linkcolor=magenta]{hyperref}
\usepackage[markup=blue, authormarkupposition=left]{changes}
\usepackage[english]{babel}
\usepackage{url}
\usepackage{siunitx}
\DeclareSIUnit \dbc {dBc}
\usepackage{soul}
\usepackage{changes}
\usepackage{cancel}
\usepackage{times}

\newcommand{\Fref}[1]{Figure~\ref{#1}}

\newcommand{\SiN}[0]{Si$_3$N$_4$~}
\newcommand{\LN}[0]{LiNbO$_3$~}

\newcommand{\SiO}[0]{SiO$_2$~}

\begin{document}
	
	\title{Ultrafast tunable photonic integrated Pockels extended-DBR laser}

	\author{Anat Siddharth}
	\email[]{authors have contributed equally}
	\affiliation{Laboratory of Photonics and Quantum Measurements, Swiss Federal Institute of Technology Lausanne (EPFL), CH-1015 Lausanne, Switzerland}
	\affiliation{Center of Quantum Science and Engineering, EPFL, CH-1015 Lausanne, Switzerland}
	
	\author{Simone Bianconi}
	\email[]{authors have contributed equally}
	\affiliation{Laboratory of Photonics and Quantum Measurements, Swiss Federal Institute of Technology Lausanne (EPFL), CH-1015 Lausanne, Switzerland}
	\affiliation{Center of Quantum Science and Engineering, EPFL, CH-1015 Lausanne, Switzerland}
	
	\author{Rui Ning Wang}
	\affiliation{Luxtelligence SA, 1025 St-Sulpice, Switzerland}
	
	\author{Zheru Qiu}
	\affiliation{Laboratory of Photonics and Quantum Measurements, Swiss Federal Institute of Technology Lausanne (EPFL), CH-1015 Lausanne, Switzerland}
	\affiliation{Center of Quantum Science and Engineering, EPFL, CH-1015 Lausanne, Switzerland}
		
	\author{Andrey S. Voloshin}
	\affiliation{Laboratory of Photonics and Quantum Measurements, Swiss Federal Institute of Technology Lausanne (EPFL), CH-1015 Lausanne, Switzerland}
	\affiliation{Center of Quantum Science and Engineering, EPFL, CH-1015 Lausanne, Switzerland}
	
	\author{Mohammad J. Bereyhi}
	\affiliation{Luxtelligence SA, 1025 St-Sulpice, Switzerland}
	
	\author{Johann Riemensberger}
	\email[]{johann.riemensberger@ntnu.no}
	\affiliation{Laboratory of Photonics and Quantum Measurements, Swiss Federal Institute of Technology Lausanne (EPFL), CH-1015 Lausanne, Switzerland}
	\affiliation{Center of Quantum Science and Engineering, EPFL, CH-1015 Lausanne, Switzerland}
	\affiliation{Norwegian University of Science and Technology, 7491 Trondheim, Norway}
	
	\author{Tobias J. Kippenberg}
	\email[]{tobias.kippenberg@epfl.ch}
	\affiliation{Laboratory of Photonics and Quantum Measurements, Swiss Federal Institute of Technology Lausanne (EPFL), CH-1015 Lausanne, Switzerland}
	\affiliation{Center of Quantum Science and Engineering, EPFL, CH-1015 Lausanne, Switzerland}
	
	\medskip
	\maketitle
	
	\noindent\textbf{Frequency-agile lasers that can simultaneously feature low noise characteristics as well as fast mode-hop-free frequency tuning are keystone components for applications ranging from frequency modulated continuous wave (FMCW) LiDAR \cite{Behroozpour:17}, to coherent optical communication \cite{klotzkin2020introduction} and gas sensing \cite{cassidy1982atmospheric}. 
	The hybrid integration of III-V gain media with low-loss photonic integrated circuits (PICs) has recently enabled integrated lasers with faster tuning and lower phase noise than the best legacy systems, including fiber lasers.
	In addition, lithium niobate on insulator (LNOI) PICs have enabled to exploit the Pockels effect to demonstrate self-injection locked hybrid lasers with tuning rates reaching peta-hertz per second \cite{snigirev2023ultrafast}. However, Pockels-tunable laser archetypes relying on high-Q optical microresonators have thus far only achieved limited output powers, are difficult to operate and stabilize due to the dynamics of self-injection locking, and require many analog control parameters. 	
	Here, we overcome this challenge by leveraging an extended distributed Bragg reflector (E-DBR) architecture to demonstrate a simple and turn-key operable frequency-agile Pockels laser that can be controlled with single analog operation and modulation inputs.
	Our laser supports a continuous mode-hop-free tuning range of over 10~GHz with good linearity and flat actuation bandwidth up to 10 MHz, while achieving over 15~mW in-fiber output power at 1545~nm, a combination unmet by legacy bulk lasers. This hybrid laser design combines an inexpensive reflective semiconductor optical amplifier (RSOA) with an electro-optic DBR PIC manufactured at wafer-scale on a LNOI platform \cite{li2023high}. This E-DBR Pockels laser features a kHz-level intrinsic linewidth and a tuning rate of 10 PHz/s with a tuning efficiency of over 550 MHz/V. We showcase the performance and flexibility of this laser in proof-of-concept coherent optical ranging (FMCW LiDAR) demonstration, achieving a 4 cm distance resolution and in a hydrogen cyanide spectroscopy experiment.
	}
	
	\begin{figure*}[htbp!]
		\centering
		\includegraphics[width=1\linewidth]{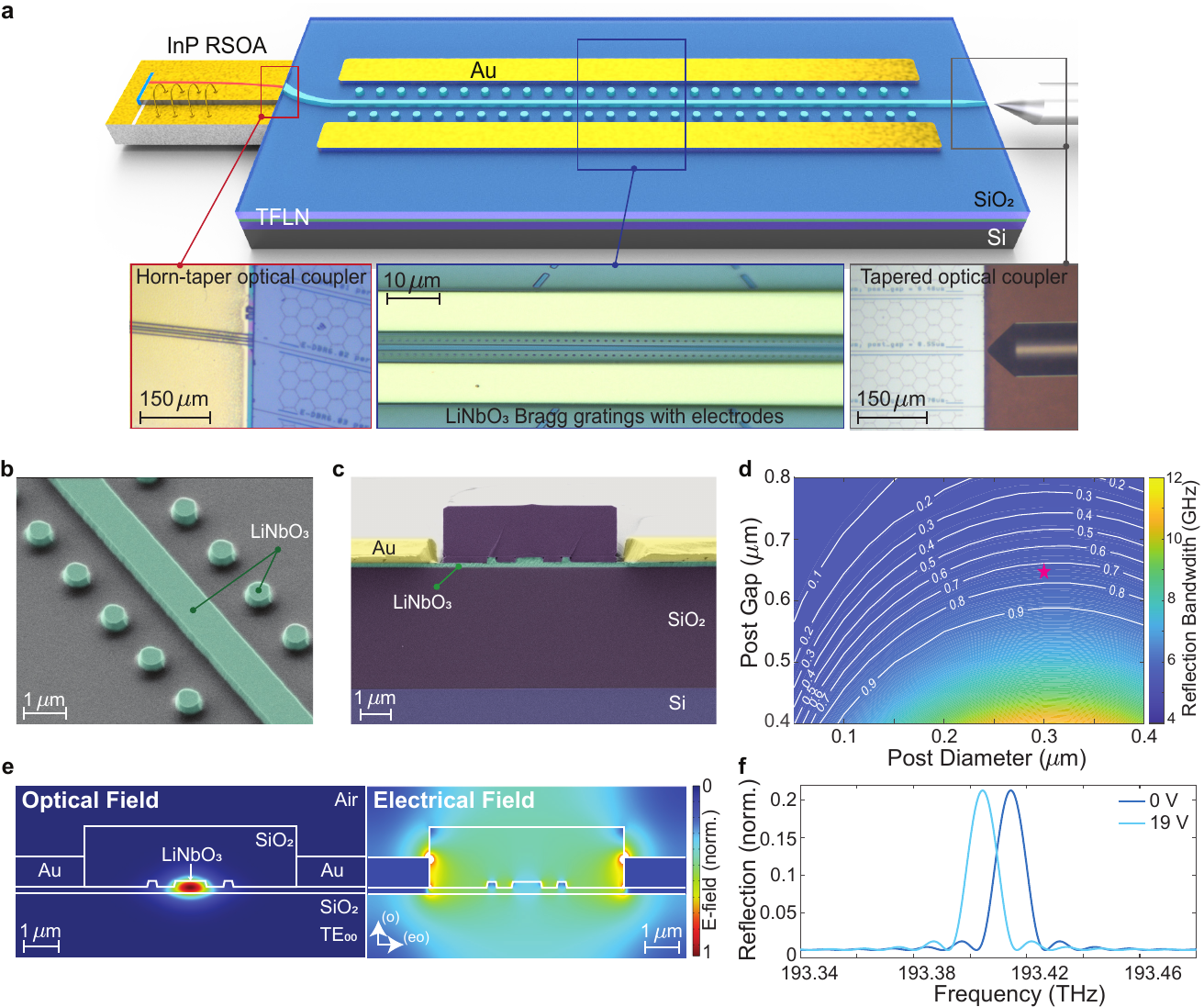}
		\caption{\textbf{Conceptual design of the hybrid integrated Pockels extended-DBR laser.}
			(a) Schematic illustration of the hybrid integrated extended-DBR (E-DBR) laser. RSOA:  reflective semiconductor optical amplifier, TFLN: thin-film lithium niobate. The insets show detailed microscope images of the butt-coupling interface between the InP RSOA and the PIC, the Bragg grating and tuning electrodes, and the double-layer taper and lensed fiber for output coupling, respectively.
			(b) Scanning electron micrograph of a detail of the \LN E-DBR PIC after the etching step to define waveguide and Bragg grating posts.
			(c) Scanning electron micrograph of the vertical cross section of the \LN photonic chip showing embedded gold electrodes and the silica (\SiO) top cladding.
			(d) Simulated normalized Bragg reflection (lines) and FWHM bandwidth of the Bragg reflection peak (colorbar) of the \LN E-DBR PIC as a function of the size and position (gap) of the grating posts.
			(e) FEM simulation of the optical and RF electric field distribution in the monolithic X-cut \LN and \SiO cladding. The RF field is normalized so that the voltage across the two electrodes is 1 V.
			(f) Simulated tuning of the Bragg grating reflection peak at applied voltages of 0 V and 19 V, depicting tuning efficiency greater than 0.5 GHz/V.
			}
		\label{fig1}
	\end{figure*}
	
	
\section{Introduction}
	Thin-film lithium niobate (\LN) integrated photonics have facilitated a manifold of novel and high performance photonic integrated circuit (PIC) technologies, owing to the fast, efficient, and low-loss refractive index modulation enabled by the strong Pockels effect and tight optical confinement \cite{zhu2021integrated}.	
	These features are not only attractive for electro-optical modulators \cite{wang2018integrated,he2019high} and transceivers but also opened new avenues to endow hybrid integrated lasers with novel ultrafast frequency modulation capabilities. 
	The self-injection locking of DFB laser diodes to Pockels tunable optical microresonators was first explored for bulk whispering gallery microresonators \cite{savchenkov2010voltage}.
	Optical microresonators based on thin-film \LN have enabled the demonstration of self-injection locked \cite{snigirev2023ultrafast,li2023high} and Vernier filter-based hybrid and heterogeneous integrated lasers \cite{Li2022,wanghigh} with frequency tuning rates reaching beyond petahertz-per-second and facilitated intracavity second harmonic generation \cite{Li2022}. 
	These frequency-agile yet low noise lasers \cite{lihachev2022low} are important building blocks for many applications ranging from frequency-modulated continuous-wave LiDAR \cite{Behroozpour:17, riemensberger2020massively}, to fast wavelength switching in telecommunications \cite{guan2018widely}, and applications in distributed fiber sensing \cite{lu2019distributed}.	
	The currently demonstrated Pockels lasers are hindered by their reliance on low-loss optical microresonators, which presents substantial fabrication challenges and complexity. Additionally, their operation requires very precise tuning and control of up to six analog parameters.
	Moreover, the utilization of high-Q microresonators has limited the mode-hop-free tuning ranges to less than 2 GHz, which is inadequate for many critical applications, particularly high-resolution atmospheric gas sensing and coherent laser ranging.
	
	Here, we present a turn-key operable frequency-agile laser based on an external distributed Bragg waveguide grating reflector (E-DBR) fabricated in a wafer-scale thin-film \LN platform. In this hybrid laser architecture, a reflective semiconductor optical amplifier (RSOA) with a high-reflectivity backside facet is employed as gain medium, where the laser mode is coupled through a low-reflectivity angled front facet onto the E-DBR PIC which constitutes the second mirror of the hybrid laser cavity. 
	E-DBR lasers with feedback circuits implemented in silicon nitride photonic integrated circuits demonstrated significant progress in recent years \cite{Xiang2021, Belt:14, Huang:19,tran2019tutorial} and their integration with piezoelectric tuning circuits has extended their application potential \cite{siddharth2024piezoelectrically}.
	The implementation of long Bragg gratings with weak coupling in thin-film \LN allows us to leverage the strong and ultrafast Pockels effect for modulation, which significantly increases the frequency tuning efficiency and reduces power consumption, scaling the tuning range from 1 GHz to 10 GHz.
	The Pockels E-DBR laser presented here is optimized to simultaneously achieve a high output power, large tuning efficiency, good linearity, and low frequency and intensity noise characteristics, surpassing the tuning performance and coherence length of many state-of-the-art monolithic semiconductor lasers, yet retaining the same simple operation routines.
	In addition, the employment of RSOA as a gain medium instead of more complex DFB laser chips allows to significantly reduce the cost of the laser components, which together with the commercial wafer-scale thin film \LN platform establishes a pathway to cost-effective and wafer-scale volume manufacturing of the chip modules of these hybrid Pockels lasers which is expected to be highly advantageous in applications such as frequency-modulated LiDAR \cite{riemensberger2020massively,lihachev2022low}, anemometry \cite{feneyrou2017frequency}, optical communications \cite{klotzkin2020introduction}, gas sensing \cite{dykema2023feasibility} and metrology \cite{udem2002optical}.
	In this simple, two element hybrid laser cavity the lasing frequency is selected and tuned by the reflection peak of the DBR structure and the optical phase accumulated in the RSOA controlled by the drive current. The non-resonant nature of RSOA facilitates locked operation over a very wide span of parameters and does not require the alignment of two independent laser or microcring cavities omitting the need for precise temperature control and microheater-based phase shifters. 
	As such, the \LN PIC acts as a tunable intracavity wavelength-selective element, which can be effectively tuned via the Pockels effect, ensuring continuous and mode-hop free frequency tuning, and making the laser operation turn-key.	
	
	\begin{figure*}[htbp!]
		\centering
		\includegraphics[width=1\linewidth]{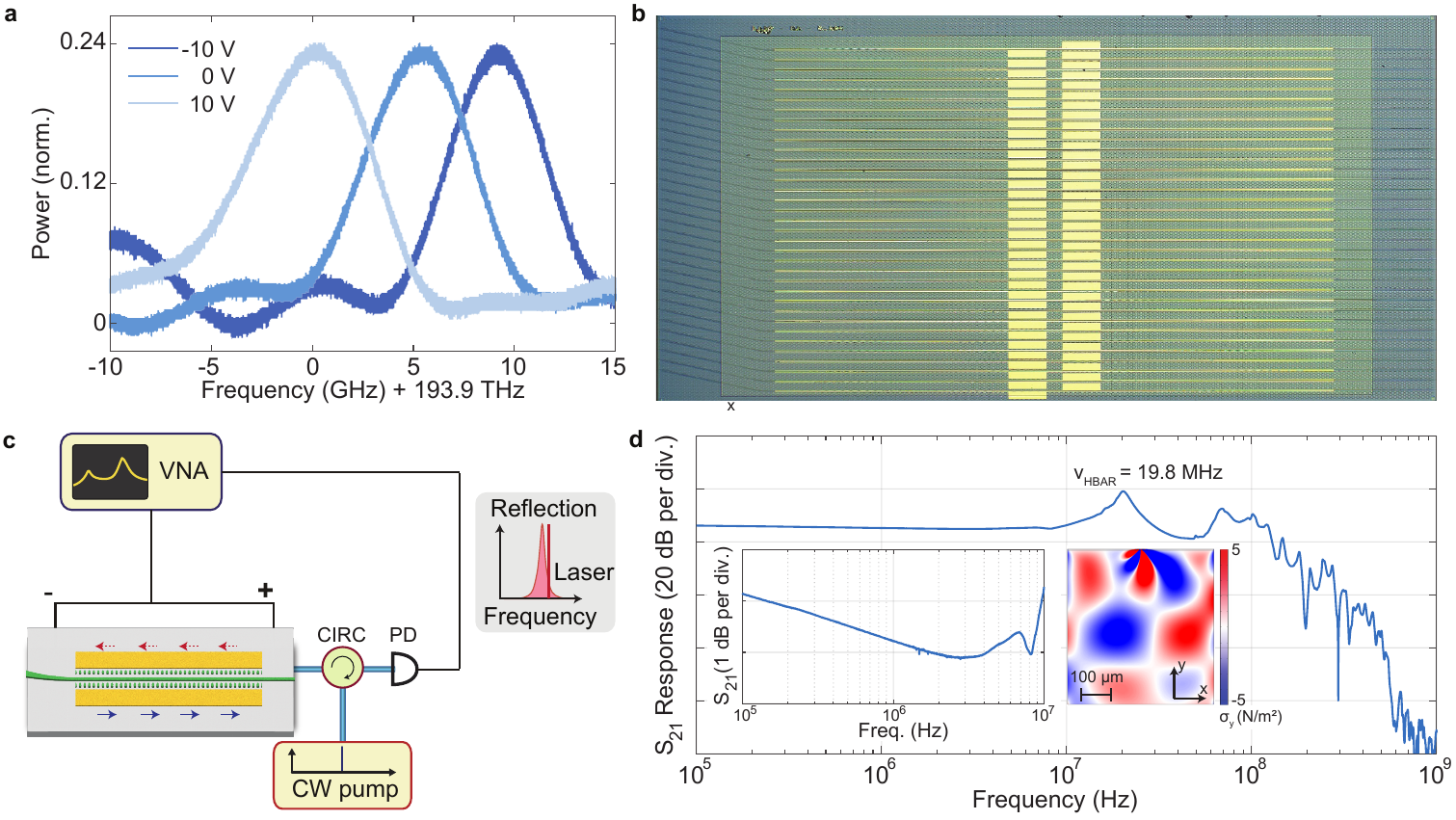}
		\caption{\textbf{Characterization of the \LN Bragg grating.}
			(a) Voltage tuning of the center frequency of the Bragg reflection from the DBR PIC with tuning efficiency of 550 MHz/V.
			(b) Photograph of the photonic chip (\texttt{D148\_01\_F2\_C3}) with Au electrodes, comprising the 7.25 mm-long DBR.
			(c) Measurement setup for the response characterization of electro-optic tuning. An external cavity laser (ECDL) is tuned to the side of the grating reflection peak, the reflected light is isolated using an optical circulator (CIRC), and the S$_{21}$ tuning response is characterized with a vector network analyzer (VNA).
			(d) Electro-optic S$_{21}$ tuning response of the DBR grating up to 1~GHz. The left inset shows a zoomed-in part of the spectrum, depicting a flat actuation response up to 10 MHz with a deviation of 1 dB over 10 MHz. The right inset shows the vertical stress $\sigma_y$ distribution of the fundamental acoustic resonant mode at 19.8 MHz. 		
			}
		\label{fig2}
	\end{figure*}

\section{E-DBR Design and Characterization}
	The architecture of the \LN PIC-based E-DBR laser is depicted in \Fref{fig1}a: an indium phosphide (InP) RSOA is butt-coupled to the PIC containing two taper couplers and the E-DBR grating flanked by gold electrodes. 
	The PIC is fabricated using a wafer-scale process on a 400 nm-thick X-cut single-crystalline thin-film \LN and contains 35 waveguide grating devices on a chip 5$\times$10~mm in area. The DBR grating is formed by etching small \LN posts on either side of the 1~$\mu$m-wide waveguide ridge, as shown in \Fref{fig1}b and in the PIC cross section in \Fref{fig1}c. Both the waveguides and the Bragg grating posts are defined by etching structures for a depth of 200~nm using a diamond-like carbon etch mask \cite{li2023high}, and leaving a slab of 200~nm (see Methods). 
	This relatively thin film and shallow etch is chosen to ensure operation in the single mode regime, as well as to facilitate the coupling between the center waveguide and the Bragg gratings posts on either side.
	Light is coupled from the RSOA to the photonic chip through an angled waveguide taper to minimize reflection at the interface while optimizing the power coupling, and the laser output is coupled to a lensed optical fiber through a second straight taper. 
	Both the PIC input and output waveguides utilize a bi-layer taper architecture \cite{he2019low} (cf. Supplementary Notes), which was optimized to achieve a coupling loss below 1.5 dB per facet. 	

	The Bragg grating structure is 7.25~mm long with a period of 1.27979 $\mu$m, forming a $3^\mathrm{rd}$ order grating with 300~nm wide posts placed at a distance of 630~nm from the waveguide top edge. 
	Since the grating is designed for $3^\mathrm{rd}$ order reflection, we used spatial Fourier analysis to isolate the m=3 coefficient determining the coupling strength $\kappa$, which determines the peak reflectivity and bandwidth of the grating (cf. Supplementary Notes). The geometry of the Bragg grating was optimized by using 2D finite element simulations to extract the effective index modulation as a function of the posts dimension and their position relative to the waveguide, as shown in \Fref{fig1}d: the 300~nm wide posts placed at a distance of 630~nm from the waveguide edge yields an estimated peak reflectivity around 75~\% and a full-width half maximum reflection peak bandwidth of 6.5~GHz. The grating was designed with a uniform coupling strength along its length, i.e. without apodization \cite{ashry2014investigating}. 	
	For Pockels tuning of the laser, 900~nm-thick gold (Au) electrodes are deposited on either side of the Bragg grating structure, 3 $\mu$m away from the waveguide ridge, as shown in the PIC cross section in \Fref{fig1}c. 
	\Fref{fig2}e displays the simulated optical waveguide mode and the electric field distribution between the electrodes: due to the large RF permittivity of \LN, the electric field in the \LN ridge is stronger in the cladding \SiO. As a result, the electrodes are placed in direct contact with \LN film in order to maximize the actuation efficiency and minimize the bias drift towards low frequencies, observed in \SiO-buffered electrodes \cite{salvestrini2011analysis}. 
	The electrodes are placed conservatively at a total gap of 7~$\mu$m to minimize optical losses and not interfere with the evanescent field in the grating posts and avoid optical losses, as detailed in Supplementary notes.

	\Fref{fig2}a shows the DC tuning of the DBR reflection peak over 10~GHz with only 20~V of applied voltage, matching the simulated voltage-length product (V$_{\pi}$L) of 4~V$\cdot$cm. These reflection measurements were performed by coupling the output of an external cavity laser (ECDL) through a lensed fiber from the PIC output port (on the right edge of the PIC in \Fref{fig2}b), and scanning its wavelength across the Bragg gratings reflection peaks, while applying different voltages at the electrodes. Calibrated reflection and transmission measurements shown in Supplementary notes yielded an off-chip reflectivity of 30~\% and a FWHM bandwidth of 8~GHz, with a sidelobe at +8 GHz offset. The presence of a close-in side lobe indicates a local distortion of the grating Bragg period, possibly due to height variations of the \LN thin film. 
	Additionally we observe reflection side lobes are observed that are likely stemming from the lack of grating apodization in our laser design and the parasitic reflection from the straight output facet. The measured reflection wavelength of 1545.9~nm only deviates by 1.2\% from the design.

	We further characterized the electrical actuation bandwidth of the Pockels grating as depicted in \Fref{fig2}c (see Methods). We found a very uniform tuning response up to a frequency of 10~MHz (\Fref{fig2}d) with a deviation of 1 dB over 10 MHz. The broad peak at 19.8~MHz originates from the piezoelectric effect driving the fundamental bulk acoustic resonance similar to earlier observations in \SiN-based lasers \cite{lihachev2023frequency,tian2020hybrid}. At higher frequencies the modulation response drops due to the bandwidth limitations of the DC needle probes. 	
		
	\begin{figure*}[htbp!]
		\centering
		\includegraphics[width=1\linewidth]{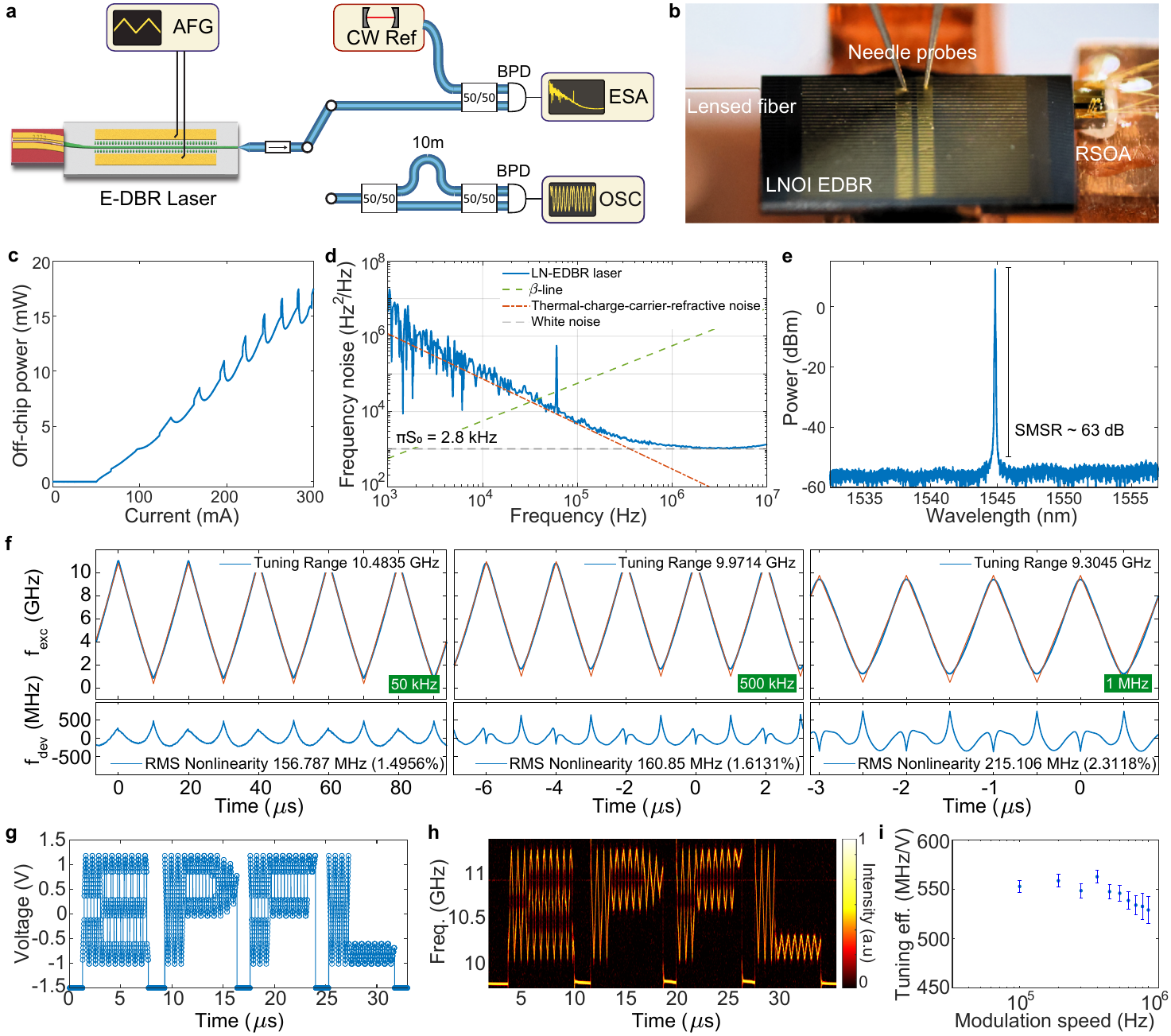}
		\caption{\textbf{Pockels E-DBR laser operation.}
			(a) Schematic of the experimental setup used to characterize the laser power, frequency noise and tuning. The laser current is set to achieve single mode operation and the laser frequency is modulated via an arbitrary waveform generator (AFG). The laser frequency noise is measured by measuring on a balanced photodetector (BPD) the heterodyne beat note with an ECDL stabilized and filtered using a bulk cavity, recorded and analyzed on an electrical spectrum analyzer (ESA). The laser tuning is characterized by measuring the homodyne beat note through a 10~m imbalanced Mach-Zehnder interferometer (iMZI), recorded on a digital oscilloscope (OSC).
			(b) Photograph of the \LN DBR PIC on the coupling stage, butt-coupled RSOA, output lensed fiber and needs probes for applying electric field across the electrodes.			
			(c) E-DBR laser output power as function of RSOA drive current. The laser set points are chosen at a local minima of curve, maximizing mode-hop free laser tuning range.
			(d) Single-sided laser frequency noise power spectral density $S_{\nu}(f)$. The peak at 70~kHz stems from a mechanical mode of the overhanging PIC. 
			(e) Optical spectrum of E-DBR laser featuring a side-mode suppression ratio (SMSR) of 63~dB at a resolution bandwidth of 0.02~nm.
			(f) Characterization of hybrid laser tuning range, speed and linearity. The E-DBR laser is modulated with triangular waveforms at speeds between 10 kHz (left) and 1~MHz (right). The nonlinearity is determined by fitting an ideal triangular waveform to the tuning curve obtained via Hilbert transform of the iMZI output.
			(g) Voltage pattern applied to the electrodes from an arbitrary waveform generator, resembling the EPFL logo.
			(h) Time-frequency spectrogram showing laser frequency evolution in the form of the EPFL logo, reaching a tuning rate of 3 PHz/s.
			(i) Tuning efficiency of the Pockels E-DBR laser at different modulation speeds, i.e. the repetition rate of the the triangular voltage ramp applied on the electrodes. 
			}
		\label{fig3}
	\end{figure*}

\section{E-DBR laser operation, tunability, and noise}
	The hybrid laser cavity is assembled by butt-coupling an InP RSOA to the facet of the thin-film \LN PIC to inject light into the optical waveguide taper.  
	We scanned the drive current of the RSOA to determine the operation points of the laser. 
	Due to the thermo-optic and plasma-dispersion effects in the RSOA gain section, tuning the drive current will induce an optical phase shift in the gain waveguide effectively tuning the intracavity phase of the hybrid integrated laser, which alleviates the need for an additional thermal phase shifter \cite{baveja2010self}. 
	Hence, the hybrid integrated E-DBR laser may be operated with a single analog control parameter, i.e. the drive current, and a single analog tuning parameter, i.e. the electro-optic modulation voltage.	
	The experimental setup to measure the laser frequency noise and tuning linearity is depicted in \Fref{fig3}a and a photograph is added in \Fref{fig3}b. 

	The frequency noise power spectral density $S_\nu(f)$ was extracted the beatnote signal between the E-DBR laser and a bulk cavity stabilized and filtered external cavity tunable diode laser (cf. \Fref{fig3}d). 
	We found a frequency noise level of 800~Hz$^2$/Hz at an offset frequency of 2~MHz, corresponding to an intrinsic linewidth of 2.8~kHz (see Methods). 
	At lower offset frequencies the frequency noise is increased with a slope of $f^{-(1.2-1.5)}$, which is likely limited by thermal-charge-carrier-refractive noise \cite{zhang2023fundamental}. 
	The FWHM optical linewidth for an integration time of 1~ms is 191.5~kHz. 
	We also observed a distinct peak in the noise spectrum at 70~kHz that we attribute to an undamped mechanical flapping mode of the 10~mm $\times$ 5~mm photonic chip. 
	The optical spectrum of the laser was recorded with a resolution bandwidth of 0.02~nm and features a very high side mode suppression ratio (SMSR) of 63~dB, corresponding to 56~dB at a resolution bandwidth of 0.1~nm, at a fiber-coupled output power of 12.5~mW (\Fref{fig3}e).
	
	Next, we measured the laser frequency tuning linearity and efficiency. 
	Motivated by the potential application of our laser for long-range and high precision frequency-modulated continuous wave LiDAR, we apply triangular waveforms to the electrodes with modulation frequencies between 10~kHz and 1~MHz and an amplitude of 17.5~V$_{pp}$. 
	The tuning efficiency of the Pockels E-DBR laser is 550~MHz/V at a modulation speed of 100~kHz of the triangular voltage ramp. This efficiency decreases to 525~MHz/V at a modulation speed of 1 MHz, as shown in \Fref{fig3}i. 
	
	We recorded the frequency modulation at the output of an imbalanced Mach-Zehnder Interferometer (iMZI) and extract the frequency modulation using a Hilbert transform and phase extraction algorithm \cite{ahn2007analysis}. 
	We fitted the extracted frequency modulation with an ideal triangular signal and plot the deviation below the signal and fit. 
	Modulating at frequencies of 10 kHz 100 kHz, and 1 MHz, we found a tuning range of 10.48~GHz, 9.97~GHz, and 9.30~GHz with a relative root-mean square (RMS) nonlinearity of 1.4\%, 1.6\%, and 2.3\%, respectively.
	We also measure the residual amplitude modulation (RAM) of the Pockels E-DBR laser while modulating it at a frequency of 100 kHz with a 17.5~V$_{pp}$ amplitude. 
	We also measured the residual amplitude modulation (RAM) of the Pockels E-DBR laser while modulating it at a frequency of 100 kHz with a 17.5~V$_{pp}$ amplitude (cf. Supplementary Notes). The integrated power spectral density of the RAM up to 100 MHz was found to be -35.94 dB.
	
	Another advantage is that the \LN E-DBR does not require resonant excitation of the modulating element yet achieves high efficiency and fast tuning, hence we can modulate the laser in an arbitrary fashion while maintaining high tuning speed. 
	Using an arbitrary waveform generator, we programmed it to reproduce the EPFL logo, as shown in \Fref{fig3}g, and applied this voltage signal to the \LN DBR PIC. The laser frequency evolution was measured by the iMZI, and the result of the time–frequency analysis is shown in \Fref{fig3}h. 
	The analysis of the arbitary tuning pattern reveals a tuning rate in excess of 3 PHz/s that increases to 9.3 PHz/s in the case of linear triangular actuation. We note that the tuning speed is not limited by the performance of the RSOA or the grating, but by the probe electrodes and voltage amplifier.

\section{Optical coherent ranging}
	\begin{figure*}[htbp!]
		\centering
		\includegraphics[width=1\linewidth]{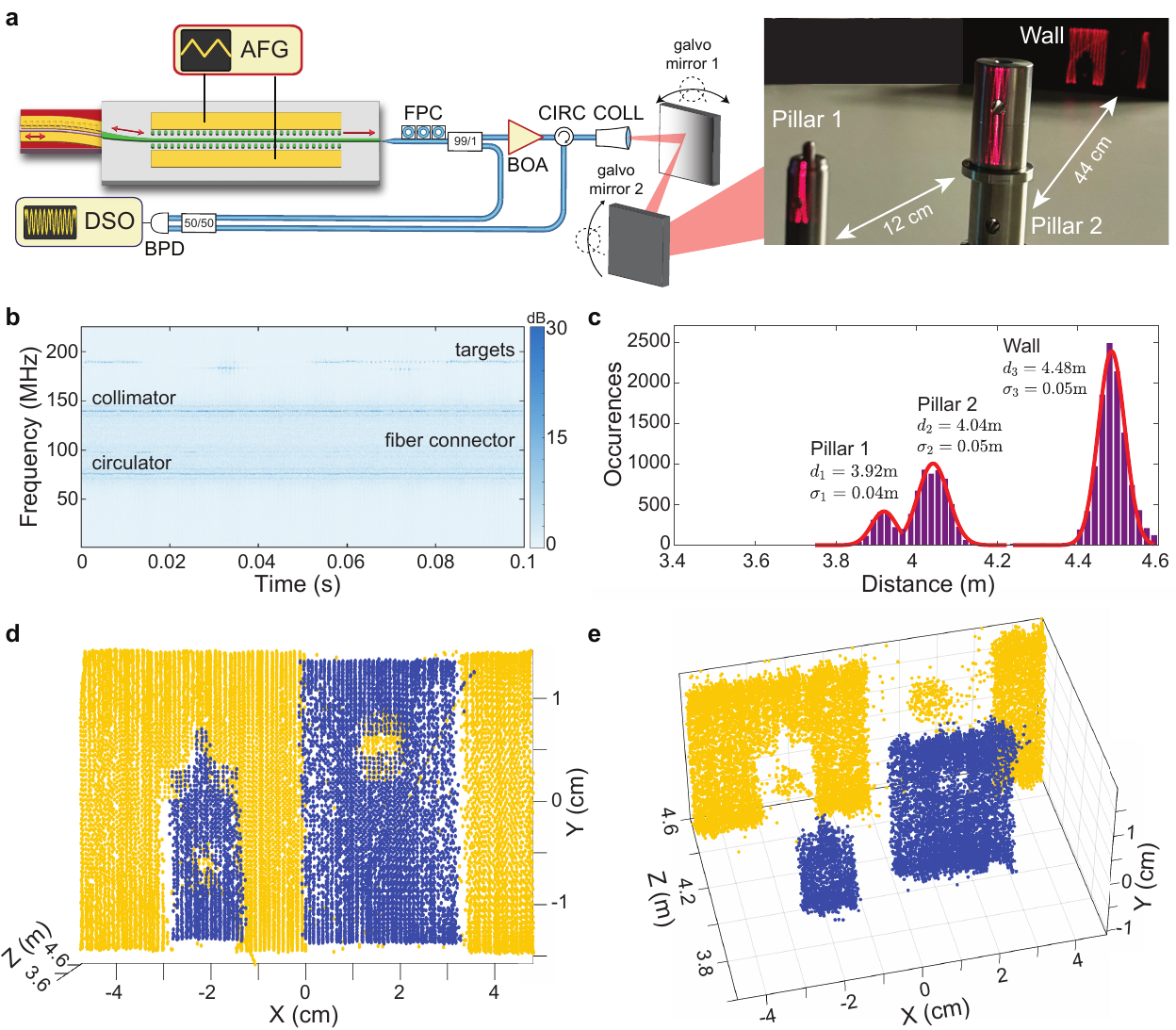}
		\caption{\textbf{Proof-of-concept demonstration of FMCW LiDAR}
			(a) Schematic of the experimental setup for coherent optical ranging based on frequency-modulated continuous wave (FMCW) LiDAR with a photograph of the target scene - two metal posts of 1~mm and 2.5~mm diameter, and a flat wall behind. DSO: digital sampling oscilloscope, AFG: arbitrary waveform generator, FPC: fiber polarization controller, EDFA: erbium-doped fiber amplifier, CIRC: circulator, COL: collimator.
			(b) Time-frequency spectrogram of the signal from the target, showing the reflection peaks from the collimator, fiber connector and circulator together with the reflection peaks from the target scene.
			(c) Histogram showing the distribution of the calculated range values of the three target objects: the average distance values and their standard deviations are reported for each target.
			(d,e) Point-cloud representations of the measured target scene from different perspectives, obtained using a beam scanning pattern with $1$~kHz vertical and $10$~Hz horizontal triangular scanning frequencies.       			
			}
		\label{fig4}
	\end{figure*}
	
	Frequency agility is a key requirement for long-range frequency-modulated continuous wave (FMCW) LiDAR, enabling fast, linear, and hysteresis-free tuning. 
	The ranging signal is inferred from the beat frequency of a laser source with triangular-shaped frequency modulation that is reflected from the target and detected by mixing it with the original signal on a fast photodiode. 
	The FMCW range resolution is determined by the tuning range as \cite{riemensberger2020massively}:
	\begin{equation}
			\Delta x = \dfrac{c}{2B},
	\end{equation}
	where $B$ is the bandwidth of the linear frequency chirp. 
	We used the fast tunability of the Pockels E-DBR laser to conduct a proof-of-concept optical coherent ranging experiment in a lab environment using the FMCW LiDAR scheme. 
	The coherent ranging setup is depicted in \Fref{fig4}a: the transmitted and return signals are both coupled through the same collimator, paired with a circulator for directional isolation.
	We used two pedestal pillar posts with diameters of 1~mm and 2.5~mm, placed on a table 4 meters away from the collimator. 
	The distance between the pillar posts is around 10~cm, and the wall is around 40~cm away from the second pillar. A photograph of the target scene and the beam-scanning pattern are shown in \Fref{fig4}a.
	
	The time-frequency spectrograms contain 20,000 time slices with a typical SNR of 20 dB for the target. 
	The beat note signals at 75 MHz, 100 MHz, and 140 MHz in \Fref{fig4}b are due to the reflections from the circulator, fiber connector, and collimator respectively, which are typical of a mono-static LiDAR setup \cite{snigirev2023ultrafast,lihachev2022low}. 
	The signals observed between 150 and 200~MHz correspond to the target scene. For each time slice, the measured beat note is linearized and filtered as described in Methods, to select the frequency peak corresponding to the target reflection. 
	Each of these frequency peaks is converted into range information (i.e. radial coordinate) using calibrated tuning and ranging information (see Methods). The fundamental depth resolution of the measurement is determined by the 10-GHz tuning range of the laser, resulting in around 1.5 cm \cite{riemensberger2020massively}.
	\Fref{fig4}c shows the distribution of detected target ranges: the cluster at a distance of 3.92 m corresponds to the first pillar post, 4.04 m to the second pillar post, and 4.48 m to the back wall, matching the measured target distances. \Fref{fig4}d and e show the resulting point cloud representations of the scene with distance-based coloring: the pillar posts are mostly visible in blue, and the wall behind in yellow.
	This proof-of-concept coherent ranging demonstration showcase the Pockels laser performance and ease of operation in LiDAR applications, where the combination of good range resolution (less than 5~cm) with simplicity of the laser operation and ranging setup can prove highly advantageous in many applications requiring low size, weight and power (SWaP).

\section{Packaged Pockels E-DBR laser and HCN spectroscopy}
	\begin{figure*}[htb!]
	\centering
	\includegraphics[width=1\linewidth]{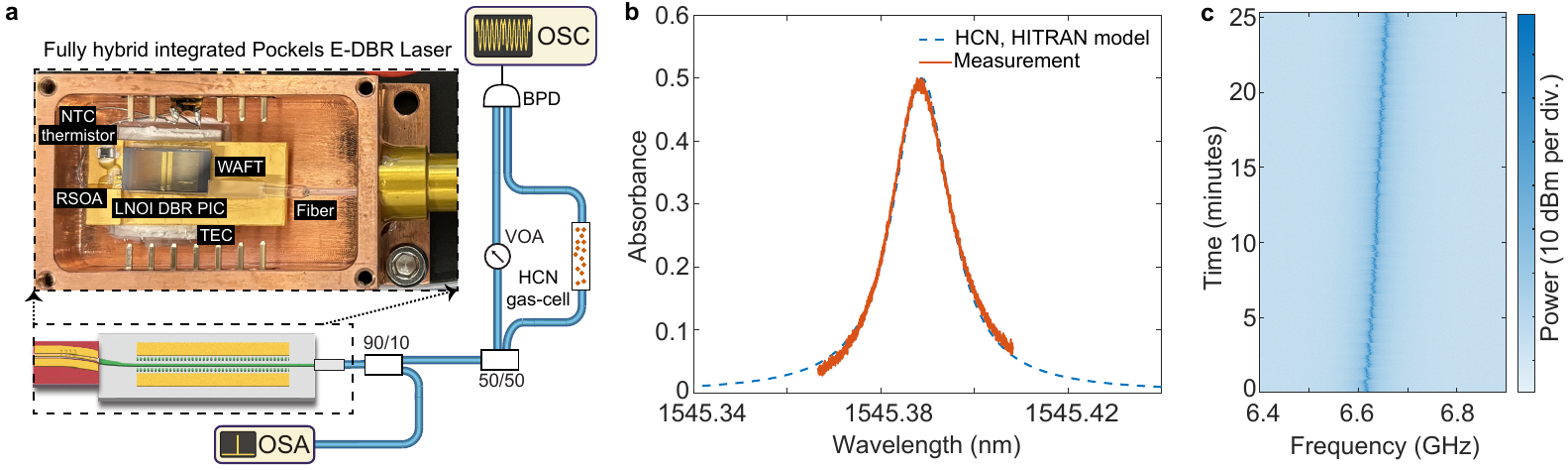}
	 \caption{\textbf{Proof-of-concept demonstration of Hydrogen Cyanide (HCN) spectroscopy}
	   (a) Schematic of the experimental setup for HCN gas sensing based on a packaged photonic integrated Pockels E-DBR laser. The inset shows the Pockels laser housed in a custom butterfly mount with an integrated thermo-electric cooler (TEC), thermistor (NTC) and coupled using a waveguide array to fiber transposer (WAFT). OSA: optical spectrum analyzer, OSC: oscilloscope, VOA: variable optical attenuator, BPD: balanced photodetector.
	   (b) Absorption spectrum of HCN around 1545 nm measured with the Pockels E-DBR laser, showing good agreement with the simulated absorbance from the HITRAN database.
	   (c) Time-frequency spectrogram of the fully hybrid integrated Pockels E-DBR laser, illustrating the long-term stability of the laser.             
	 }
	 \label{fig5}
	\end{figure*}
	Fast tunable low-noise lasers are highly sought after for spectroscopic and sensing applications, where the frequency agility and narrow linewidth of these lasers allows to precisely target an absorption feature of a species of interest to infer its concentration. Besides fast tuning capabilities, low residual amplitude modulation (RAM) and good chirp linearity are also crucial for spectroscopic applications.
	
	For this purpose, we packaged the photonic integrated laser in a custom butterfly mount with an integrated thermo-electric cooler (TEC), by using epoxy and active alignment to glue the RSOA chip, photonic chip and a waveguide array to fiber transposer (WAFT) together as depicted in \Fref{fig5}a. We observed a substantial reduction in RAM in the packaged laser as compared to the prototype laser (cf. Supplementary Notes). We attribute this to a reduction in electrostatic noise granted by the physical shielding and wire-bonding of the on-chip electrodes as compared to the open probing configuration shown in Figure \ref{fig3}. Moreover, the packaged laser features the capability of controlling the temperature of the PIC using a TEC, which allows to tune the Bragg grating reflection over several nm to target specific spectral absorption lines.
	
	In order to showcase the advantageous performance of such packaged laser, we performed a spectroscopic measurement of an absorption line of Hydrogen Cyanide (HCN) at around 1545 nm. We used a fiber-coupled absorption cell (Wavelength Reference) with 20 Torr HCN natural isotopic abundance, 16.5 cm path length, and a calibrated optical spectrum analyzer to target the appropriate wavelength corresponding to the target absorption feature. We tuned the center wavelength of the laser over several nm using the TEC integrated in the butterfly package and then used the on-chip electrodes to electro-optically chirp the laser wavelength over 40 pm across the targeted absorption feature of HCN, yielding the absorption trace shown in Figure \ref{fig5}b. This proof-of-concept measurement also offers further validation of the tuning range of the laser, by comparing the width of the measured absorption feature to the feature simulated from the HITRAN database \cite{rothman2021history}, showing good agreement, as shown in Figure \ref{fig5}b.
	
\section{Discussion and Outlook}
	 
	We demonstrate a Pockels-tunable hybrid integrated laser. This hybrid laser design integrates a cost-effective reflective semiconductor optical amplifier (RSOA) with an electro-optic DBR PIC, fabricated at wafer scale on a lithium niobate on insulator (LNOI) platform. 
	We use a new design for high-order and slab-coupled Bragg post gratings on \LN that allows to generate weakly coupled gratings with 10~GHz reflection bandwidths and millimeter coherence lengths. 
	Our design approach alleviates the need for unreliable sub-nm crenelation of the slanted waveguide side-walls and short grating periods ensuring manufacturability with contemporary PIC foundry processes. 
	The \LN post grating can function as an ultrafast tunable mirror through its strong frequency modulation via the Pockels effect	 thereby endowing a hybrid integrated laser with continuous and mode-hop-free frequency agility. 

	Based on this grating design, we build a turn-key operable frequency-agile Pockels laser, capable of large electro-optical continuous mode-hop free frequency modulation. 
	The Pockels-tunable E-DBR features a more straightforward and reliable laser operation, distinguishing it from other frequency-agile lasers that require precise control of multiple parameters. 
	Additionally, using an RSOA as the gain medium, instead of more complex distributed feedback (DFB) laser chips reduces component costs compared to earlier demonstrations of Pockels-tunable lasers using both integrated photonics \cite{snigirev2023ultrafast,li2023high}, and bulk whispering gallery microresonators \cite{savchenkov2010voltage}. 
	We show a mode-hop free tuning range of 10~GHz, modulation speed of 1~MHz, residual nonlinearity of 1~\%, fiber-coupled output power of 15~mW, and frequency noise level of 800~Hz$^2$/Hz, corresponding 2.8~kHz intrinsic (lorentzian) linewidth and a tuning efficiency exceeding 550~MHz/V. 
	We achieve a tuning rate of 10~PHz/s, which is limited not by the laser architecture but the available electronics. 
	Our extensive numerical simulations of the hybrid laser system predicts that we can increase the tuning range beyond 20~GHz and reduce the nonlinearity further to the 0.1~\% level by apodizing the grating and angling the output facet to spoil spurious reflections.  
	 
	In summary, our advancements highlight the potential for wafer-scale volume manufacturing of high-performance hybrid Pockels lasers, making them  suitable for various applications. The laser's performance can further be improved by co-integrating the DBR with a low-loss, high-Q microresonator \cite{siddharth2024piezoelectrically}, yet at the cost of increased complexity.  
    The performance and versatility of this laser have been demonstrated through proof-of-concept experiments, achieving below 5 cm distance resolution in coherent optical ranging (FMCW LiDAR) and conducting hydrogen cyanide spectroscopy. 
    The transparency range of \LN in principle supports hybrid laser operation into the mid-infrared frequency range. 
    These proof-of-concept measurements showcasing the laser tuning range and linearity as well as the stability and simplicity of operation, demonstrate the potential of such lasers in spectroscopy, and will encourage further development of the technology towards such applications.

\section*{Methods}
	
\subsection{Device fabrication}
	We used 4-inch lithium niobate on insulator wafers with 400~nm X-cut lithium niobate thin-film on 4.7~$\mu$m \SiO. The photonic circuit designs were exposed using a DUV stepper photolithography tool and etched into lithium niobate by ion-beam etching using a diamond-like carbon etch-mask \cite{li2023high}. 
	The lithium niobate thin-film is etched to a depth of 200~nm, creating ridge waveguides and leaving a lithium niobate slab over the rest of the wafer. 
	In a second photolithography step, tapered waveguides are etched into the lithium niobate slab at the edges with a tapering tip width of 400~nm of the dies to reduce optical coupling losses. 
	At this point the lithium niobate structures are clad with 2~$\mu$m of high-density \SiO using ICP-CVD deposition. By judicious application of RF substrate biasing during deposition, the circuits can be clad without leaving any voids in the gaps between densely packed photonic structures, e.g. the gaps between the grating posts and the center waveguide.
	Finally, in a third photolithography, the cladding is etched down to the lithium niobate slab, and 900~nm thick gold electrodes are fabricated by a lift-off process inside these recesses in the cladding \cite{li2023high}.
	
\subsection{Characterization of the electro-optic S$_{21}$ response of the DBR PIC}
	To measure the electro-optic S$_{21}$ response of the DBR PIC, we tuned an ECDL laser (Toptica CTL 1550nm ) to the slope of the grating reflection peak. This setup allowed the conversion of small signal modulation into a detectable response due to the rapid change in reflection efficiency with wavelength at this operating point. A small signal from a vector network analyzer (VNA) was applied, and the resulting modulated signal was directed through a circulator to a fast photodiode. The photodiode detected the intensity-modulated signal, which was then analyzed by the VNA to obtain the S$_{21}$ parameter. The S$_{21}$ parameter provided a detailed frequency domain characterization of the electro-optic response of the Bragg gratings.

\subsection{Characterization of the laser linewidth}
	We measure the laser frequency noise power spectral density using heterodyne beat spectroscopy with an external-cavity diode laser (Toptica CTL 1550 nm) that is locked to and filtered by a free-space Fabry-P\'erot (FP) cavity with a free spectral range (FSR) of 1.5GHz and a linewidth of 70~kHz. 
	The beatnote of the two signals was detected on a fast photodiode, and its electrical output was sent to an electrical spectrum analyzer (Rohde \& Schwarz FSVA3030). 
	The beat note signal was down-mixed with our electrical spectrum analyzer (ESA) recording the in-phase and quadrature signals, which were digitized with the 40~MHz internal digitizer of the ESA. 
	The recorded data for the in-phase and quadrature components of the beatnote were processed using Welch’s method \cite{welch1967use} to retrieve the single-sided phase noise power spectral density ($S_{\phi\phi}$), which was converted to frequency noise ($S_{ff}$) using $S_{ff} = f^2 \times S_{\phi\phi}$. 
	To estimate the integrated linewidth, we integrated the frequency noise spectra from the intersection of the power spectral density with the beta-line down to the integration time of the measurement \cite{di2010simple}. 
	The area under the curve \( A \) is then recalculated to provide the full-width at half-maximum (FWHM) measure of the linewidth using $\text{FWHM} = \sqrt{8 \ln(2) \cdot A}$.

\subsection{Coherent ranging experiment and signal data processing}
	For the coherent ranging experiment, the RSOA drive current was set to 194~mA for an output power of 8.2~mW. The laser output is amplified to around 20 mW using a booster optical amplifier (Thorlabs BOA-59855), with a portion of the laser output tapped before amplification and used as LO in delayed homodyne detection. The laser beam is scanned across the target scene using two galvo mirrors (Thorlabs GVS112), which are driven by triangular signals at 1 kHz and 10 Hz for the vertical and horizontal directions, respectively. Polar and azimuthal coordinates were obtained from the galvo scanner mirrors driving signals, which were recorded on the same oscilloscope.
	The test scene consisted of two stainless steel metal pedestal pillar posts (Thorlabs) with 3~mm  and 4~mm holes, respectively. 	
	The data for the point cloud were collected within the total time interval of 100~ms.
	In order to calibrate the frequency excursion during the ranging measurements, 5\% of the laser output was sent to a reference iMZI fiber interferometer of calibrated length. Based on the calculated frequency excursion, an ideal range resolution of 1.5~cm is inferred. The recorded return signal was divided into 20,000 segment, one per each ramp segment, and subjected to digital signal processing in order to locate the scene elements in space. First, the signal recorded from the reference iMZI was linearized using Hilbert transform, and then used to resample the return signal to improve the SNR. Second, the zero-padded short-time Fourier transforms of the beatnote oscillograms from the target and the reference iMZI were evaluated. Third, the maxima in each spectrogram was selected by excluding the frequency peaks corresponding to the circulator, fiber connector and collimator shown in \Fref{fig4}b. Finally, we converted the frequency points to the distance domain using the calibrated frequency excursion inferred from the iMZI, and subtracted the distance from the laser to the collimator so that the point-cloud distance is given with respect to the collimator aperture position.

\begin{footnotesize}
		

		\noindent \textbf{Author Contributions}:
		A.S. and S.B. did the experiments and analyzed the data with the help of Z.Q and J.R.
		J.R. simulated and designed the devices. 
		R.N.W. and M.J.B. fabricated the device. 
		A.S. and J.R. characterized the devices.
		S.B. and A.V. packaged the device. 
		A.S., J.R., and S.B. wrote the manuscript with input from all authors. 
		T.J.K. and J.R. supervised the project.
				
		\noindent \textbf{Funding Information and Disclaimer}: This publication was supported by Contract W911NF2120248 (NINJA LASER) from the Defense Advanced Research Projects Agency (DARPA), Microsystems Technology Office (MTO), as well as the Swiss National Science Foundation (SNSF) through grant number 211728 (BRIDGE) and grant number 221171 (SPARK - MetPIL).
		A.S. acknowledges support from the European Space Technology Centre with ESA Contract No. 4000135357/21/NL/GLC/my. 
		J.R. acknowledges funding from the SNSF Ambizione fellowship (No. 201923) and Onsager Fellowship from NTNU.
		
		\noindent \textbf{Disclosures}: The authors declare no competing financial interests. T.J.K. is a co-founder and shareholder of DEEPLIGHT SA, a startup commercializing PIC-based frequency-agile, low noise lasers.
				
		\noindent \textbf{Data Availability Statement}: The code and data used to produce the plots within this work will be released on the repository \texttt{Zenodo} upon publication of this preprint.
		
		\noindent\textbf{Correspondence and requests for materials} should be addressed to J.R.
	\end{footnotesize}
\bibliography{citations}

\begin{thebibliography}{36}%
\makeatletter
\providecommand \@ifxundefined [1]{%
 \@ifx{#1\undefined}
}%
\providecommand \@ifnum [1]{%
 \ifnum #1\expandafter \@firstoftwo
 \else \expandafter \@secondoftwo
 \fi
}%
\providecommand \@ifx [1]{%
 \ifx #1\expandafter \@firstoftwo
 \else \expandafter \@secondoftwo
 \fi
}%
\providecommand \natexlab [1]{#1}%
\providecommand \enquote  [1]{``#1''}%
\providecommand \bibnamefont  [1]{#1}%
\providecommand \bibfnamefont [1]{#1}%
\providecommand \citenamefont [1]{#1}%
\providecommand \href@noop [0]{\@secondoftwo}%
\providecommand \href [0]{\begingroup \@sanitize@url \@href}%
\providecommand \@href[1]{\@@startlink{#1}\@@href}%
\providecommand \@@href[1]{\endgroup#1\@@endlink}%
\providecommand \@sanitize@url [0]{\catcode `\\12\catcode `\$12\catcode
  `\&12\catcode `\#12\catcode `\^12\catcode `\_12\catcode `\%12\relax}%
\providecommand \@@startlink[1]{}%
\providecommand \@@endlink[0]{}%
\providecommand \url  [0]{\begingroup\@sanitize@url \@url }%
\providecommand \@url [1]{\endgroup\@href {#1}{\urlprefix }}%
\providecommand \urlprefix  [0]{URL }%
\providecommand \Eprint [0]{\href }%
\providecommand \doibase [0]{https://doi.org/}%
\providecommand \selectlanguage [0]{\@gobble}%
\providecommand \bibinfo  [0]{\@secondoftwo}%
\providecommand \bibfield  [0]{\@secondoftwo}%
\providecommand \translation [1]{[#1]}%
\providecommand \BibitemOpen [0]{}%
\providecommand \bibitemStop [0]{}%
\providecommand \bibitemNoStop [0]{.\EOS\space}%
\providecommand \EOS [0]{\spacefactor3000\relax}%
\providecommand \BibitemShut  [1]{\csname bibitem#1\endcsname}%
\let\auto@bib@innerbib\@empty
\bibitem [{\citenamefont {Behroozpour}\ \emph {et~al.}(2017)\citenamefont
  {Behroozpour}, \citenamefont {Sandborn}, \citenamefont {Quack}, \citenamefont
  {Seok}, \citenamefont {Matsui}, \citenamefont {Wu},\ and\ \citenamefont
  {Boser}}]{Behroozpour:17}%
  \BibitemOpen
  \bibfield  {author} {\bibinfo {author} {\bibfnamefont {B.}~\bibnamefont
  {Behroozpour}}, \bibinfo {author} {\bibfnamefont {P.~A.~M.}\ \bibnamefont
  {Sandborn}}, \bibinfo {author} {\bibfnamefont {N.}~\bibnamefont {Quack}},
  \bibinfo {author} {\bibfnamefont {T.-J.}\ \bibnamefont {Seok}}, \bibinfo
  {author} {\bibfnamefont {Y.}~\bibnamefont {Matsui}}, \bibinfo {author}
  {\bibfnamefont {M.~C.}\ \bibnamefont {Wu}},\ and\ \bibinfo {author}
  {\bibfnamefont {B.~E.}\ \bibnamefont {Boser}},\ }\bibfield  {title} {\bibinfo
  {title} {Electronic-photonic integrated circuit for 3d microimaging},\ }\href
  {https://doi.org/10.1109/JSSC.2016.2621755} {\bibfield  {journal} {\bibinfo
  {journal} {IEEE Journal of Solid-State Circuits}\ }\textbf {\bibinfo {volume}
  {52}},\ \bibinfo {pages} {161} (\bibinfo {year} {2017})}\BibitemShut
  {NoStop}%
\bibitem [{\citenamefont {Klotzkin}(2020)}]{klotzkin2020introduction}%
  \BibitemOpen
  \bibfield  {author} {\bibinfo {author} {\bibfnamefont {D.~J.}\ \bibnamefont
  {Klotzkin}},\ }\href
  {https://link.springer.com/book/10.1007/978-3-030-24501-6} {\emph {\bibinfo
  {title} {Introduction to semiconductor lasers for optical communications}}}\
  (\bibinfo  {publisher} {Springer},\ \bibinfo {year} {2020})\BibitemShut
  {NoStop}%
\bibitem [{\citenamefont {Cassidy}\ and\ \citenamefont
  {Reid}(1982)}]{cassidy1982atmospheric}%
  \BibitemOpen
  \bibfield  {author} {\bibinfo {author} {\bibfnamefont {D.~T.}\ \bibnamefont
  {Cassidy}}\ and\ \bibinfo {author} {\bibfnamefont {J.}~\bibnamefont {Reid}},\
  }\bibfield  {title} {\bibinfo {title} {Atmospheric pressure monitoring of
  trace gases using tunable diode lasers},\ }\href
  {https://opg.optica.org/ao/fulltext.cfm?uri=ao-21-7-1185&id=25681} {\bibfield
   {journal} {\bibinfo  {journal} {Applied Optics}\ }\textbf {\bibinfo {volume}
  {21}},\ \bibinfo {pages} {1185} (\bibinfo {year} {1982})}\BibitemShut
  {NoStop}%
\bibitem [{\citenamefont {Snigirev}\ \emph {et~al.}(2023)\citenamefont
  {Snigirev}, \citenamefont {Riedhauser}, \citenamefont {Lihachev},
  \citenamefont {Churaev}, \citenamefont {Riemensberger}, \citenamefont {Wang},
  \citenamefont {Siddharth}, \citenamefont {Huang}, \citenamefont {M{\"o}hl},
  \citenamefont {Popoff} \emph {et~al.}}]{snigirev2023ultrafast}%
  \BibitemOpen
  \bibfield  {author} {\bibinfo {author} {\bibfnamefont {V.}~\bibnamefont
  {Snigirev}}, \bibinfo {author} {\bibfnamefont {A.}~\bibnamefont
  {Riedhauser}}, \bibinfo {author} {\bibfnamefont {G.}~\bibnamefont
  {Lihachev}}, \bibinfo {author} {\bibfnamefont {M.}~\bibnamefont {Churaev}},
  \bibinfo {author} {\bibfnamefont {J.}~\bibnamefont {Riemensberger}}, \bibinfo
  {author} {\bibfnamefont {R.~N.}\ \bibnamefont {Wang}}, \bibinfo {author}
  {\bibfnamefont {A.}~\bibnamefont {Siddharth}}, \bibinfo {author}
  {\bibfnamefont {G.}~\bibnamefont {Huang}}, \bibinfo {author} {\bibfnamefont
  {C.}~\bibnamefont {M{\"o}hl}}, \bibinfo {author} {\bibfnamefont
  {Y.}~\bibnamefont {Popoff}}, \emph {et~al.},\ }\bibfield  {title} {\bibinfo
  {title} {Ultrafast tunable lasers using lithium niobate integrated
  photonics},\ }\href {https://www.nature.com/articles/s41586-023-05724-2}
  {\bibfield  {journal} {\bibinfo  {journal} {Nature}\ }\textbf {\bibinfo
  {volume} {615}},\ \bibinfo {pages} {411} (\bibinfo {year}
  {2023})}\BibitemShut {NoStop}%
\bibitem [{\citenamefont {Li}\ \emph {et~al.}(2023)\citenamefont {Li},
  \citenamefont {Wang}, \citenamefont {Lihachev}, \citenamefont {Zhang},
  \citenamefont {Tan}, \citenamefont {Churaev}, \citenamefont {Kuznetsov},
  \citenamefont {Siddharth}, \citenamefont {Bereyhi}, \citenamefont
  {Riemensberger} \emph {et~al.}}]{li2023high}%
  \BibitemOpen
  \bibfield  {author} {\bibinfo {author} {\bibfnamefont {Z.}~\bibnamefont
  {Li}}, \bibinfo {author} {\bibfnamefont {R.~N.}\ \bibnamefont {Wang}},
  \bibinfo {author} {\bibfnamefont {G.}~\bibnamefont {Lihachev}}, \bibinfo
  {author} {\bibfnamefont {J.}~\bibnamefont {Zhang}}, \bibinfo {author}
  {\bibfnamefont {Z.}~\bibnamefont {Tan}}, \bibinfo {author} {\bibfnamefont
  {M.}~\bibnamefont {Churaev}}, \bibinfo {author} {\bibfnamefont
  {N.}~\bibnamefont {Kuznetsov}}, \bibinfo {author} {\bibfnamefont
  {A.}~\bibnamefont {Siddharth}}, \bibinfo {author} {\bibfnamefont {M.~J.}\
  \bibnamefont {Bereyhi}}, \bibinfo {author} {\bibfnamefont {J.}~\bibnamefont
  {Riemensberger}}, \emph {et~al.},\ }\bibfield  {title} {\bibinfo {title}
  {High density lithium niobate photonic integrated circuits},\ }\href
  {https://www.nature.com/articles/s41467-023-40502-8} {\bibfield  {journal}
  {\bibinfo  {journal} {Nature Communications}\ }\textbf {\bibinfo {volume}
  {14}},\ \bibinfo {pages} {4856} (\bibinfo {year} {2023})}\BibitemShut
  {NoStop}%
\bibitem [{\citenamefont {Zhu}\ \emph {et~al.}(2021)\citenamefont {Zhu},
  \citenamefont {Shao}, \citenamefont {Yu}, \citenamefont {Cheng},
  \citenamefont {Desiatov}, \citenamefont {Xin}, \citenamefont {Hu},
  \citenamefont {Holzgrafe}, \citenamefont {Ghosh}, \citenamefont
  {Shams-Ansari} \emph {et~al.}}]{zhu2021integrated}%
  \BibitemOpen
  \bibfield  {author} {\bibinfo {author} {\bibfnamefont {D.}~\bibnamefont
  {Zhu}}, \bibinfo {author} {\bibfnamefont {L.}~\bibnamefont {Shao}}, \bibinfo
  {author} {\bibfnamefont {M.}~\bibnamefont {Yu}}, \bibinfo {author}
  {\bibfnamefont {R.}~\bibnamefont {Cheng}}, \bibinfo {author} {\bibfnamefont
  {B.}~\bibnamefont {Desiatov}}, \bibinfo {author} {\bibfnamefont
  {C.}~\bibnamefont {Xin}}, \bibinfo {author} {\bibfnamefont {Y.}~\bibnamefont
  {Hu}}, \bibinfo {author} {\bibfnamefont {J.}~\bibnamefont {Holzgrafe}},
  \bibinfo {author} {\bibfnamefont {S.}~\bibnamefont {Ghosh}}, \bibinfo
  {author} {\bibfnamefont {A.}~\bibnamefont {Shams-Ansari}}, \emph {et~al.},\
  }\bibfield  {title} {\bibinfo {title} {Integrated photonics on thin-film
  lithium niobate},\ }\href
  {https://opg.optica.org/aop/fulltext.cfm?uri=aop-13-2-242&id=450625}
  {\bibfield  {journal} {\bibinfo  {journal} {Advances in Optics and
  Photonics}\ }\textbf {\bibinfo {volume} {13}},\ \bibinfo {pages} {242}
  (\bibinfo {year} {2021})}\BibitemShut {NoStop}%
\bibitem [{\citenamefont {Wang}\ \emph {et~al.}(2018)\citenamefont {Wang},
  \citenamefont {Zhang}, \citenamefont {Chen}, \citenamefont {Bertrand},
  \citenamefont {Shams-Ansari}, \citenamefont {Chandrasekhar}, \citenamefont
  {Winzer},\ and\ \citenamefont {Lon{\v{c}}ar}}]{wang2018integrated}%
  \BibitemOpen
  \bibfield  {author} {\bibinfo {author} {\bibfnamefont {C.}~\bibnamefont
  {Wang}}, \bibinfo {author} {\bibfnamefont {M.}~\bibnamefont {Zhang}},
  \bibinfo {author} {\bibfnamefont {X.}~\bibnamefont {Chen}}, \bibinfo {author}
  {\bibfnamefont {M.}~\bibnamefont {Bertrand}}, \bibinfo {author}
  {\bibfnamefont {A.}~\bibnamefont {Shams-Ansari}}, \bibinfo {author}
  {\bibfnamefont {S.}~\bibnamefont {Chandrasekhar}}, \bibinfo {author}
  {\bibfnamefont {P.}~\bibnamefont {Winzer}},\ and\ \bibinfo {author}
  {\bibfnamefont {M.}~\bibnamefont {Lon{\v{c}}ar}},\ }\bibfield  {title}
  {\bibinfo {title} {Integrated lithium niobate electro-optic modulators
  operating at cmos-compatible voltages},\ }\href
  {https://www.nature.com/articles/s41586-018-0551-y} {\bibfield  {journal}
  {\bibinfo  {journal} {Nature}\ }\textbf {\bibinfo {volume} {562}},\ \bibinfo
  {pages} {101} (\bibinfo {year} {2018})}\BibitemShut {NoStop}%
\bibitem [{\citenamefont {He}\ \emph {et~al.}(2019{\natexlab{a}})\citenamefont
  {He}, \citenamefont {Xu}, \citenamefont {Ren}, \citenamefont {Jian},
  \citenamefont {Ruan}, \citenamefont {Xu}, \citenamefont {Gao}, \citenamefont
  {Sun}, \citenamefont {Wen}, \citenamefont {Zhou} \emph
  {et~al.}}]{he2019high}%
  \BibitemOpen
  \bibfield  {author} {\bibinfo {author} {\bibfnamefont {M.}~\bibnamefont
  {He}}, \bibinfo {author} {\bibfnamefont {M.}~\bibnamefont {Xu}}, \bibinfo
  {author} {\bibfnamefont {Y.}~\bibnamefont {Ren}}, \bibinfo {author}
  {\bibfnamefont {J.}~\bibnamefont {Jian}}, \bibinfo {author} {\bibfnamefont
  {Z.}~\bibnamefont {Ruan}}, \bibinfo {author} {\bibfnamefont {Y.}~\bibnamefont
  {Xu}}, \bibinfo {author} {\bibfnamefont {S.}~\bibnamefont {Gao}}, \bibinfo
  {author} {\bibfnamefont {S.}~\bibnamefont {Sun}}, \bibinfo {author}
  {\bibfnamefont {X.}~\bibnamefont {Wen}}, \bibinfo {author} {\bibfnamefont
  {L.}~\bibnamefont {Zhou}}, \emph {et~al.},\ }\bibfield  {title} {\bibinfo
  {title} {High-performance hybrid silicon and lithium niobate mach--zehnder
  modulators for 100 gbit s- 1 and beyond},\ }\href
  {https://www.nature.com/articles/s41566-019-0378-6} {\bibfield  {journal}
  {\bibinfo  {journal} {Nature photonics}\ }\textbf {\bibinfo {volume} {13}},\
  \bibinfo {pages} {359} (\bibinfo {year} {2019}{\natexlab{a}})}\BibitemShut
  {NoStop}%
\bibitem [{\citenamefont {Savchenkov}\ \emph {et~al.}(2010)\citenamefont
  {Savchenkov}, \citenamefont {Ilchenko}, \citenamefont {Liang}, \citenamefont
  {Eliyahu}, \citenamefont {Matsko}, \citenamefont {Seidel},\ and\
  \citenamefont {Maleki}}]{savchenkov2010voltage}%
  \BibitemOpen
  \bibfield  {author} {\bibinfo {author} {\bibfnamefont {A.}~\bibnamefont
  {Savchenkov}}, \bibinfo {author} {\bibfnamefont {V.}~\bibnamefont
  {Ilchenko}}, \bibinfo {author} {\bibfnamefont {W.}~\bibnamefont {Liang}},
  \bibinfo {author} {\bibfnamefont {D.}~\bibnamefont {Eliyahu}}, \bibinfo
  {author} {\bibfnamefont {A.}~\bibnamefont {Matsko}}, \bibinfo {author}
  {\bibfnamefont {D.}~\bibnamefont {Seidel}},\ and\ \bibinfo {author}
  {\bibfnamefont {L.}~\bibnamefont {Maleki}},\ }\bibfield  {title} {\bibinfo
  {title} {Voltage-controlled photonic oscillator},\ }\href
  {https://opg.optica.org/ol/fulltext.cfm?uri=ol-35-10-1572&id=199274}
  {\bibfield  {journal} {\bibinfo  {journal} {Optics letters}\ }\textbf
  {\bibinfo {volume} {35}},\ \bibinfo {pages} {1572} (\bibinfo {year}
  {2010})}\BibitemShut {NoStop}%
\bibitem [{\citenamefont {Li}\ \emph {et~al.}(2022)\citenamefont {Li},
  \citenamefont {Chang}, \citenamefont {Wu}, \citenamefont {Staffa},
  \citenamefont {Ling}, \citenamefont {Javid}, \citenamefont {Xue},
  \citenamefont {He}, \citenamefont {Lopez-rios}, \citenamefont {Morin},
  \citenamefont {Wang}, \citenamefont {Shen}, \citenamefont {Zeng},
  \citenamefont {Zhu}, \citenamefont {Vahala}, \citenamefont {Bowers},\ and\
  \citenamefont {Lin}}]{Li2022}%
  \BibitemOpen
  \bibfield  {author} {\bibinfo {author} {\bibfnamefont {M.}~\bibnamefont
  {Li}}, \bibinfo {author} {\bibfnamefont {L.}~\bibnamefont {Chang}}, \bibinfo
  {author} {\bibfnamefont {L.}~\bibnamefont {Wu}}, \bibinfo {author}
  {\bibfnamefont {J.}~\bibnamefont {Staffa}}, \bibinfo {author} {\bibfnamefont
  {J.}~\bibnamefont {Ling}}, \bibinfo {author} {\bibfnamefont {U.~A.}\
  \bibnamefont {Javid}}, \bibinfo {author} {\bibfnamefont {S.}~\bibnamefont
  {Xue}}, \bibinfo {author} {\bibfnamefont {Y.}~\bibnamefont {He}}, \bibinfo
  {author} {\bibfnamefont {R.}~\bibnamefont {Lopez-rios}}, \bibinfo {author}
  {\bibfnamefont {T.~J.}\ \bibnamefont {Morin}}, \bibinfo {author}
  {\bibfnamefont {H.}~\bibnamefont {Wang}}, \bibinfo {author} {\bibfnamefont
  {B.}~\bibnamefont {Shen}}, \bibinfo {author} {\bibfnamefont {S.}~\bibnamefont
  {Zeng}}, \bibinfo {author} {\bibfnamefont {L.}~\bibnamefont {Zhu}}, \bibinfo
  {author} {\bibfnamefont {K.~J.}\ \bibnamefont {Vahala}}, \bibinfo {author}
  {\bibfnamefont {J.~E.}\ \bibnamefont {Bowers}},\ and\ \bibinfo {author}
  {\bibfnamefont {Q.}~\bibnamefont {Lin}},\ }\bibfield  {title} {\bibinfo
  {title} {Integrated pockels laser},\ }\href
  {https://doi.org/10.1038/s41467-022-33101-6} {\bibfield  {journal} {\bibinfo
  {journal} {Nature Communications}\ }\textbf {\bibinfo {volume} {13}},\
  \bibinfo {pages} {5344} (\bibinfo {year} {2022})}\BibitemShut {NoStop}%
\bibitem [{\citenamefont {Wang}\ \emph {et~al.}()\citenamefont {Wang},
  \citenamefont {Lin}, \citenamefont {Wang}, \citenamefont {Zhang},
  \citenamefont {Ma},\ and\ \citenamefont {Cai}}]{wanghigh}%
  \BibitemOpen
  \bibfield  {author} {\bibinfo {author} {\bibfnamefont {S.}~\bibnamefont
  {Wang}}, \bibinfo {author} {\bibfnamefont {Z.}~\bibnamefont {Lin}}, \bibinfo
  {author} {\bibfnamefont {Q.}~\bibnamefont {Wang}}, \bibinfo {author}
  {\bibfnamefont {X.}~\bibnamefont {Zhang}}, \bibinfo {author} {\bibfnamefont
  {R.}~\bibnamefont {Ma}},\ and\ \bibinfo {author} {\bibfnamefont
  {X.}~\bibnamefont {Cai}},\ }\bibfield  {title} {\bibinfo {title}
  {High-performance integrated laser based on thin-film lithium niobate
  photonics for coherent ranging},\ }\href
  {https://onlinelibrary.wiley.com/doi/10.1002/lpor.202400224} {\bibinfo
  {journal} {Laser \& Photonics Reviews}\ ,\ \bibinfo {pages}
  {2400224}}\BibitemShut {NoStop}%
\bibitem [{\citenamefont {Lihachev}\ \emph {et~al.}(2022)\citenamefont
  {Lihachev}, \citenamefont {Riemensberger}, \citenamefont {Weng},
  \citenamefont {Liu}, \citenamefont {Tian}, \citenamefont {Siddharth},
  \citenamefont {Snigirev}, \citenamefont {Shadymov}, \citenamefont {Voloshin},
  \citenamefont {Wang} \emph {et~al.}}]{lihachev2022low}%
  \BibitemOpen
\bibfield  {journal} {  }\bibfield  {author} {\bibinfo {author} {\bibfnamefont
  {G.}~\bibnamefont {Lihachev}}, \bibinfo {author} {\bibfnamefont
  {J.}~\bibnamefont {Riemensberger}}, \bibinfo {author} {\bibfnamefont
  {W.}~\bibnamefont {Weng}}, \bibinfo {author} {\bibfnamefont {J.}~\bibnamefont
  {Liu}}, \bibinfo {author} {\bibfnamefont {H.}~\bibnamefont {Tian}}, \bibinfo
  {author} {\bibfnamefont {A.}~\bibnamefont {Siddharth}}, \bibinfo {author}
  {\bibfnamefont {V.}~\bibnamefont {Snigirev}}, \bibinfo {author}
  {\bibfnamefont {V.}~\bibnamefont {Shadymov}}, \bibinfo {author}
  {\bibfnamefont {A.}~\bibnamefont {Voloshin}}, \bibinfo {author}
  {\bibfnamefont {R.~N.}\ \bibnamefont {Wang}}, \emph {et~al.},\ }\bibfield
  {title} {\bibinfo {title} {Low-noise frequency-agile photonic integrated
  lasers for coherent ranging},\ }\href
  {https://www.nature.com/articles/s41467-022-30911-6} {\bibfield  {journal}
  {\bibinfo  {journal} {Nature communications}\ }\textbf {\bibinfo {volume}
  {13}},\ \bibinfo {pages} {1} (\bibinfo {year} {2022})}\BibitemShut {NoStop}%
\bibitem [{\citenamefont {Riemensberger}\ \emph {et~al.}(2020)\citenamefont
  {Riemensberger}, \citenamefont {Lukashchuk}, \citenamefont {Karpov},
  \citenamefont {Weng}, \citenamefont {Lucas}, \citenamefont {Liu},\ and\
  \citenamefont {Kippenberg}}]{riemensberger2020massively}%
  \BibitemOpen
  \bibfield  {author} {\bibinfo {author} {\bibfnamefont {J.}~\bibnamefont
  {Riemensberger}}, \bibinfo {author} {\bibfnamefont {A.}~\bibnamefont
  {Lukashchuk}}, \bibinfo {author} {\bibfnamefont {M.}~\bibnamefont {Karpov}},
  \bibinfo {author} {\bibfnamefont {W.}~\bibnamefont {Weng}}, \bibinfo {author}
  {\bibfnamefont {E.}~\bibnamefont {Lucas}}, \bibinfo {author} {\bibfnamefont
  {J.}~\bibnamefont {Liu}},\ and\ \bibinfo {author} {\bibfnamefont {T.~J.}\
  \bibnamefont {Kippenberg}},\ }\bibfield  {title} {\bibinfo {title} {Massively
  parallel coherent laser ranging using a soliton microcomb},\ }\href
  {https://www.nature.com/articles/s41586-020-2239-3} {\bibfield  {journal}
  {\bibinfo  {journal} {Nature}\ }\textbf {\bibinfo {volume} {581}},\ \bibinfo
  {pages} {164} (\bibinfo {year} {2020})}\BibitemShut {NoStop}%
\bibitem [{\citenamefont {Guan}\ \emph {et~al.}(2018)\citenamefont {Guan},
  \citenamefont {Novack}, \citenamefont {Galfsky}, \citenamefont {Ma},
  \citenamefont {Fathololoumi}, \citenamefont {Horth}, \citenamefont {Huynh},
  \citenamefont {Roman}, \citenamefont {Shi}, \citenamefont {Caverley} \emph
  {et~al.}}]{guan2018widely}%
  \BibitemOpen
  \bibfield  {author} {\bibinfo {author} {\bibfnamefont {H.}~\bibnamefont
  {Guan}}, \bibinfo {author} {\bibfnamefont {A.}~\bibnamefont {Novack}},
  \bibinfo {author} {\bibfnamefont {T.}~\bibnamefont {Galfsky}}, \bibinfo
  {author} {\bibfnamefont {Y.}~\bibnamefont {Ma}}, \bibinfo {author}
  {\bibfnamefont {S.}~\bibnamefont {Fathololoumi}}, \bibinfo {author}
  {\bibfnamefont {A.}~\bibnamefont {Horth}}, \bibinfo {author} {\bibfnamefont
  {T.~N.}\ \bibnamefont {Huynh}}, \bibinfo {author} {\bibfnamefont
  {J.}~\bibnamefont {Roman}}, \bibinfo {author} {\bibfnamefont
  {R.}~\bibnamefont {Shi}}, \bibinfo {author} {\bibfnamefont {M.}~\bibnamefont
  {Caverley}}, \emph {et~al.},\ }\bibfield  {title} {\bibinfo {title}
  {Widely-tunable, narrow-linewidth iii-v/silicon hybrid external-cavity laser
  for coherent communication},\ }\href
  {https://opg.optica.org/oe/fulltext.cfm?uri=oe-26-7-7920&id=383841}
  {\bibfield  {journal} {\bibinfo  {journal} {Optics express}\ }\textbf
  {\bibinfo {volume} {26}},\ \bibinfo {pages} {7920} (\bibinfo {year}
  {2018})}\BibitemShut {NoStop}%
\bibitem [{\citenamefont {Lu}\ \emph {et~al.}(2019)\citenamefont {Lu},
  \citenamefont {Lalam}, \citenamefont {Badar}, \citenamefont {Liu},
  \citenamefont {Chorpening}, \citenamefont {Buric},\ and\ \citenamefont
  {Ohodnicki}}]{lu2019distributed}%
  \BibitemOpen
  \bibfield  {author} {\bibinfo {author} {\bibfnamefont {P.}~\bibnamefont
  {Lu}}, \bibinfo {author} {\bibfnamefont {N.}~\bibnamefont {Lalam}}, \bibinfo
  {author} {\bibfnamefont {M.}~\bibnamefont {Badar}}, \bibinfo {author}
  {\bibfnamefont {B.}~\bibnamefont {Liu}}, \bibinfo {author} {\bibfnamefont
  {B.~T.}\ \bibnamefont {Chorpening}}, \bibinfo {author} {\bibfnamefont
  {M.~P.}\ \bibnamefont {Buric}},\ and\ \bibinfo {author} {\bibfnamefont
  {P.~R.}\ \bibnamefont {Ohodnicki}},\ }\bibfield  {title} {\bibinfo {title}
  {Distributed optical fiber sensing: Review and perspective},\ }\href
  {https://pubs.aip.org/aip/apr/article/6/4/041302/124295/Distributed-optical-fiber-sensing-Review-and}
  {\bibfield  {journal} {\bibinfo  {journal} {Applied Physics Reviews}\
  }\textbf {\bibinfo {volume} {6}} (\bibinfo {year} {2019})}\BibitemShut
  {NoStop}%
\bibitem [{\citenamefont {Xiang}\ \emph {et~al.}(2021)\citenamefont {Xiang},
  \citenamefont {Guo}, \citenamefont {Jin}, \citenamefont {Wu}, \citenamefont
  {Peters}, \citenamefont {Xie}, \citenamefont {Chang}, \citenamefont {Shen},
  \citenamefont {Wang}, \citenamefont {Yang}, \citenamefont {Kinghorn},
  \citenamefont {Paniccia}, \citenamefont {Vahala}, \citenamefont {Morton},\
  and\ \citenamefont {Bowers}}]{Xiang2021}%
  \BibitemOpen
  \bibfield  {author} {\bibinfo {author} {\bibfnamefont {C.}~\bibnamefont
  {Xiang}}, \bibinfo {author} {\bibfnamefont {J.}~\bibnamefont {Guo}}, \bibinfo
  {author} {\bibfnamefont {W.}~\bibnamefont {Jin}}, \bibinfo {author}
  {\bibfnamefont {L.}~\bibnamefont {Wu}}, \bibinfo {author} {\bibfnamefont
  {J.}~\bibnamefont {Peters}}, \bibinfo {author} {\bibfnamefont
  {W.}~\bibnamefont {Xie}}, \bibinfo {author} {\bibfnamefont {L.}~\bibnamefont
  {Chang}}, \bibinfo {author} {\bibfnamefont {B.}~\bibnamefont {Shen}},
  \bibinfo {author} {\bibfnamefont {H.}~\bibnamefont {Wang}}, \bibinfo {author}
  {\bibfnamefont {Q.-F.}\ \bibnamefont {Yang}}, \bibinfo {author}
  {\bibfnamefont {D.}~\bibnamefont {Kinghorn}}, \bibinfo {author}
  {\bibfnamefont {M.}~\bibnamefont {Paniccia}}, \bibinfo {author}
  {\bibfnamefont {K.~J.}\ \bibnamefont {Vahala}}, \bibinfo {author}
  {\bibfnamefont {P.~A.}\ \bibnamefont {Morton}},\ and\ \bibinfo {author}
  {\bibfnamefont {J.~E.}\ \bibnamefont {Bowers}},\ }\bibfield  {title}
  {\bibinfo {title} {High-performance lasers for fully integrated silicon
  nitride photonics},\ }\href {https://doi.org/10.1038/s41467-021-26804-9}
  {\bibfield  {journal} {\bibinfo  {journal} {Nature Communications}\ }\textbf
  {\bibinfo {volume} {12}},\ \bibinfo {pages} {6650} (\bibinfo {year}
  {2021})}\BibitemShut {NoStop}%
\bibitem [{\citenamefont {Belt}\ and\ \citenamefont
  {Blumenthal}(2014)}]{Belt:14}%
  \BibitemOpen
  \bibfield  {author} {\bibinfo {author} {\bibfnamefont {M.}~\bibnamefont
  {Belt}}\ and\ \bibinfo {author} {\bibfnamefont {D.~J.}\ \bibnamefont
  {Blumenthal}},\ }\bibfield  {title} {\bibinfo {title} {Erbium-doped waveguide
  dbr and dfb laser arrays integrated within an ultra-low-loss si3n4
  platform},\ }\href {https://doi.org/10.1364/OE.22.010655} {\bibfield
  {journal} {\bibinfo  {journal} {Opt. Express}\ }\textbf {\bibinfo {volume}
  {22}},\ \bibinfo {pages} {10655} (\bibinfo {year} {2014})}\BibitemShut
  {NoStop}%
\bibitem [{\citenamefont {Huang}\ \emph {et~al.}(2019)\citenamefont {Huang},
  \citenamefont {Tran}, \citenamefont {Guo}, \citenamefont {Peters},
  \citenamefont {Komljenovic}, \citenamefont {Malik}, \citenamefont {Morton},\
  and\ \citenamefont {Bowers}}]{Huang:19}%
  \BibitemOpen
  \bibfield  {author} {\bibinfo {author} {\bibfnamefont {D.}~\bibnamefont
  {Huang}}, \bibinfo {author} {\bibfnamefont {M.~A.}\ \bibnamefont {Tran}},
  \bibinfo {author} {\bibfnamefont {J.}~\bibnamefont {Guo}}, \bibinfo {author}
  {\bibfnamefont {J.}~\bibnamefont {Peters}}, \bibinfo {author} {\bibfnamefont
  {T.}~\bibnamefont {Komljenovic}}, \bibinfo {author} {\bibfnamefont
  {A.}~\bibnamefont {Malik}}, \bibinfo {author} {\bibfnamefont {P.~A.}\
  \bibnamefont {Morton}},\ and\ \bibinfo {author} {\bibfnamefont {J.~E.}\
  \bibnamefont {Bowers}},\ }\bibfield  {title} {\bibinfo {title} {High-power
  sub-khz linewidth lasers fully integrated on silicon},\ }\href
  {https://doi.org/10.1364/OPTICA.6.000745} {\bibfield  {journal} {\bibinfo
  {journal} {Optica}\ }\textbf {\bibinfo {volume} {6}},\ \bibinfo {pages} {745}
  (\bibinfo {year} {2019})}\BibitemShut {NoStop}%
\bibitem [{\citenamefont {Tran}\ \emph {et~al.}(2019)\citenamefont {Tran},
  \citenamefont {Huang},\ and\ \citenamefont {Bowers}}]{tran2019tutorial}%
  \BibitemOpen
  \bibfield  {author} {\bibinfo {author} {\bibfnamefont {M.~A.}\ \bibnamefont
  {Tran}}, \bibinfo {author} {\bibfnamefont {D.}~\bibnamefont {Huang}},\ and\
  \bibinfo {author} {\bibfnamefont {J.~E.}\ \bibnamefont {Bowers}},\ }\bibfield
   {title} {\bibinfo {title} {Tutorial on narrow linewidth tunable
  semiconductor lasers using si/iii-v heterogeneous integration},\ }\href
  {https://pubs.aip.org/aip/app/article/4/11/111101/1061823/Tutorial-on-narrow-linewidth-tunable-semiconductor}
  {\bibfield  {journal} {\bibinfo  {journal} {APL photonics}\ }\textbf
  {\bibinfo {volume} {4}},\ \bibinfo {pages} {111101} (\bibinfo {year}
  {2019})}\BibitemShut {NoStop}%
\bibitem [{\citenamefont {Siddharth}\ \emph {et~al.}(2024)\citenamefont
  {Siddharth}, \citenamefont {Attanasio}, \citenamefont {Bianconi},
  \citenamefont {Lihachev}, \citenamefont {Zhang}, \citenamefont {Qiu},
  \citenamefont {Bancora}, \citenamefont {Kenning}, \citenamefont {Wang},
  \citenamefont {Voloshin} \emph {et~al.}}]{siddharth2024piezoelectrically}%
  \BibitemOpen
  \bibfield  {author} {\bibinfo {author} {\bibfnamefont {A.}~\bibnamefont
  {Siddharth}}, \bibinfo {author} {\bibfnamefont {A.}~\bibnamefont
  {Attanasio}}, \bibinfo {author} {\bibfnamefont {S.}~\bibnamefont {Bianconi}},
  \bibinfo {author} {\bibfnamefont {G.}~\bibnamefont {Lihachev}}, \bibinfo
  {author} {\bibfnamefont {J.}~\bibnamefont {Zhang}}, \bibinfo {author}
  {\bibfnamefont {Z.}~\bibnamefont {Qiu}}, \bibinfo {author} {\bibfnamefont
  {A.}~\bibnamefont {Bancora}}, \bibinfo {author} {\bibfnamefont
  {S.}~\bibnamefont {Kenning}}, \bibinfo {author} {\bibfnamefont {R.~N.}\
  \bibnamefont {Wang}}, \bibinfo {author} {\bibfnamefont {A.~S.}\ \bibnamefont
  {Voloshin}}, \emph {et~al.},\ }\bibfield  {title} {\bibinfo {title}
  {Piezoelectrically tunable, narrow linewidth photonic integrated extended-dbr
  lasers},\ }\href
  {https://opg.optica.org/optica/fulltext.cfm?uri=optica-11-8-1062} {\bibfield
  {journal} {\bibinfo  {journal} {Optica}\ }\textbf {\bibinfo {volume} {11}},\
  \bibinfo {pages} {1062} (\bibinfo {year} {2024})}\BibitemShut {NoStop}%
\bibitem [{\citenamefont {S{\"o}chtig}(1988)}]{sochtig1988ti}%
  \BibitemOpen
  \bibfield  {author} {\bibinfo {author} {\bibfnamefont {J.}~\bibnamefont
  {S{\"o}chtig}},\ }\bibfield  {title} {\bibinfo {title} {Ti: Linbo3 stripe
  waveguide bragg reflector gratings},\ }\href@noop {} {\bibfield  {journal}
  {\bibinfo  {journal} {Electronics letters}\ }\textbf {\bibinfo {volume}
  {24}},\ \bibinfo {pages} {844} (\bibinfo {year} {1988})}\BibitemShut
  {NoStop}%
\bibitem [{\citenamefont {Becker}\ \emph {et~al.}(2000)\citenamefont {Becker},
  \citenamefont {Oesselke}, \citenamefont {Pandavenes}, \citenamefont {Ricken},
  \citenamefont {Rochhausen}, \citenamefont {Schreiber}, \citenamefont
  {Sohler}, \citenamefont {Suche}, \citenamefont {Wessel}, \citenamefont
  {Balsamo} \emph {et~al.}}]{becker2000advanced}%
  \BibitemOpen
  \bibfield  {author} {\bibinfo {author} {\bibfnamefont {C.}~\bibnamefont
  {Becker}}, \bibinfo {author} {\bibfnamefont {T.}~\bibnamefont {Oesselke}},
  \bibinfo {author} {\bibfnamefont {J.}~\bibnamefont {Pandavenes}}, \bibinfo
  {author} {\bibfnamefont {R.}~\bibnamefont {Ricken}}, \bibinfo {author}
  {\bibfnamefont {K.}~\bibnamefont {Rochhausen}}, \bibinfo {author}
  {\bibfnamefont {G.}~\bibnamefont {Schreiber}}, \bibinfo {author}
  {\bibfnamefont {W.}~\bibnamefont {Sohler}}, \bibinfo {author} {\bibfnamefont
  {H.}~\bibnamefont {Suche}}, \bibinfo {author} {\bibfnamefont
  {R.}~\bibnamefont {Wessel}}, \bibinfo {author} {\bibfnamefont
  {S.}~\bibnamefont {Balsamo}}, \emph {et~al.},\ }\bibfield  {title} {\bibinfo
  {title} {Advanced ti: Er: Linbo/sub 3/waveguide lasers},\ }\href@noop {}
  {\bibfield  {journal} {\bibinfo  {journal} {IEEE Journal of Selected Topics
  in Quantum Electronics}\ }\textbf {\bibinfo {volume} {6}},\ \bibinfo {pages}
  {101} (\bibinfo {year} {2000})}\BibitemShut {NoStop}%
\bibitem [{\citenamefont {Feneyrou}\ \emph {et~al.}(2017)\citenamefont
  {Feneyrou}, \citenamefont {Leviandier}, \citenamefont {Minet}, \citenamefont
  {Pillet}, \citenamefont {Martin}, \citenamefont {Dolfi}, \citenamefont
  {Schlotterbeck}, \citenamefont {Rondeau}, \citenamefont {Lacondemine},
  \citenamefont {Rieu} \emph {et~al.}}]{feneyrou2017frequency}%
  \BibitemOpen
  \bibfield  {author} {\bibinfo {author} {\bibfnamefont {P.}~\bibnamefont
  {Feneyrou}}, \bibinfo {author} {\bibfnamefont {L.}~\bibnamefont
  {Leviandier}}, \bibinfo {author} {\bibfnamefont {J.}~\bibnamefont {Minet}},
  \bibinfo {author} {\bibfnamefont {G.}~\bibnamefont {Pillet}}, \bibinfo
  {author} {\bibfnamefont {A.}~\bibnamefont {Martin}}, \bibinfo {author}
  {\bibfnamefont {D.}~\bibnamefont {Dolfi}}, \bibinfo {author} {\bibfnamefont
  {J.-P.}\ \bibnamefont {Schlotterbeck}}, \bibinfo {author} {\bibfnamefont
  {P.}~\bibnamefont {Rondeau}}, \bibinfo {author} {\bibfnamefont
  {X.}~\bibnamefont {Lacondemine}}, \bibinfo {author} {\bibfnamefont
  {A.}~\bibnamefont {Rieu}}, \emph {et~al.},\ }\bibfield  {title} {\bibinfo
  {title} {Frequency-modulated multifunction lidar for anemometry, range
  finding, and velocimetry—2. experimental results},\ }\href
  {https://opg.optica.org/ao/fulltext.cfm?uri=ao-56-35-9676} {\bibfield
  {journal} {\bibinfo  {journal} {Applied optics}\ }\textbf {\bibinfo {volume}
  {56}},\ \bibinfo {pages} {9676} (\bibinfo {year} {2017})}\BibitemShut
  {NoStop}%
\bibitem [{\citenamefont {Dykema}\ \emph {et~al.}(2023)\citenamefont {Dykema},
  \citenamefont {Bianconi}, \citenamefont {Mascarenhas},\ and\ \citenamefont
  {Anderson}}]{dykema2023feasibility}%
  \BibitemOpen
  \bibfield  {author} {\bibinfo {author} {\bibfnamefont {J.~A.}\ \bibnamefont
  {Dykema}}, \bibinfo {author} {\bibfnamefont {S.}~\bibnamefont {Bianconi}},
  \bibinfo {author} {\bibfnamefont {C.}~\bibnamefont {Mascarenhas}},\ and\
  \bibinfo {author} {\bibfnamefont {J.}~\bibnamefont {Anderson}},\ }\bibfield
  {title} {\bibinfo {title} {Feasibility study of a total precipitable water
  ipda lidar from a solar-powered stratospheric aircraft},\ }\href
  {https://opg.optica.org/ao/fulltext.cfm?uri=ao-62-25-6724&id=536749}
  {\bibfield  {journal} {\bibinfo  {journal} {Applied Optics}\ }\textbf
  {\bibinfo {volume} {62}},\ \bibinfo {pages} {6724} (\bibinfo {year}
  {2023})}\BibitemShut {NoStop}%
\bibitem [{\citenamefont {Udem}\ \emph {et~al.}(2002)\citenamefont {Udem},
  \citenamefont {Holzwarth},\ and\ \citenamefont
  {H{\"a}nsch}}]{udem2002optical}%
  \BibitemOpen
  \bibfield  {author} {\bibinfo {author} {\bibfnamefont {T.}~\bibnamefont
  {Udem}}, \bibinfo {author} {\bibfnamefont {R.}~\bibnamefont {Holzwarth}},\
  and\ \bibinfo {author} {\bibfnamefont {T.~W.}\ \bibnamefont {H{\"a}nsch}},\
  }\bibfield  {title} {\bibinfo {title} {Optical frequency metrology},\ }\href
  {https://www.nature.com/articles/416233a} {\bibfield  {journal} {\bibinfo
  {journal} {Nature}\ }\textbf {\bibinfo {volume} {416}},\ \bibinfo {pages}
  {233} (\bibinfo {year} {2002})}\BibitemShut {NoStop}%
\bibitem [{\citenamefont {He}\ \emph {et~al.}(2019{\natexlab{b}})\citenamefont
  {He}, \citenamefont {Zhang}, \citenamefont {Shams-Ansari}, \citenamefont
  {Zhu}, \citenamefont {Wang},\ and\ \citenamefont {Marko}}]{he2019low}%
  \BibitemOpen
  \bibfield  {author} {\bibinfo {author} {\bibfnamefont {L.}~\bibnamefont
  {He}}, \bibinfo {author} {\bibfnamefont {M.}~\bibnamefont {Zhang}}, \bibinfo
  {author} {\bibfnamefont {A.}~\bibnamefont {Shams-Ansari}}, \bibinfo {author}
  {\bibfnamefont {R.}~\bibnamefont {Zhu}}, \bibinfo {author} {\bibfnamefont
  {C.}~\bibnamefont {Wang}},\ and\ \bibinfo {author} {\bibfnamefont
  {L.}~\bibnamefont {Marko}},\ }\bibfield  {title} {\bibinfo {title} {Low-loss
  fiber-to-chip interface for lithium niobate photonic integrated circuits},\
  }\href {https://opg.optica.org/ol/fulltext.cfm?uri=ol-44-9-2314&id=409322}
  {\bibfield  {journal} {\bibinfo  {journal} {Optics letters}\ }\textbf
  {\bibinfo {volume} {44}},\ \bibinfo {pages} {2314} (\bibinfo {year}
  {2019}{\natexlab{b}})}\BibitemShut {NoStop}%
\bibitem [{\citenamefont {Ashry}\ \emph {et~al.}(2014)\citenamefont {Ashry},
  \citenamefont {Elrashidi}, \citenamefont {Mahros}, \citenamefont {Alhaddad},\
  and\ \citenamefont {Elleithy}}]{ashry2014investigating}%
  \BibitemOpen
  \bibfield  {author} {\bibinfo {author} {\bibfnamefont {I.}~\bibnamefont
  {Ashry}}, \bibinfo {author} {\bibfnamefont {A.}~\bibnamefont {Elrashidi}},
  \bibinfo {author} {\bibfnamefont {A.}~\bibnamefont {Mahros}}, \bibinfo
  {author} {\bibfnamefont {M.}~\bibnamefont {Alhaddad}},\ and\ \bibinfo
  {author} {\bibfnamefont {K.}~\bibnamefont {Elleithy}},\ }\bibfield  {title}
  {\bibinfo {title} {Investigating the performance of apodized fiber bragg
  gratings for sensing applications},\ }in\ \href
  {https://ieeexplore.ieee.org/abstract/document/6820640} {\emph {\bibinfo
  {booktitle} {Proceedings of the 2014 Zone 1 conference of the american
  society for engineering education}}}\ (\bibinfo {organization} {IEEE},\
  \bibinfo {year} {2014})\ pp.\ \bibinfo {pages} {1--5}\BibitemShut {NoStop}%
\bibitem [{\citenamefont {Salvestrini}\ \emph {et~al.}(2011)\citenamefont
  {Salvestrini}, \citenamefont {Guilbert}, \citenamefont {Fontana},
  \citenamefont {Abarkan},\ and\ \citenamefont
  {Gille}}]{salvestrini2011analysis}%
  \BibitemOpen
  \bibfield  {author} {\bibinfo {author} {\bibfnamefont {J.~P.}\ \bibnamefont
  {Salvestrini}}, \bibinfo {author} {\bibfnamefont {L.}~\bibnamefont
  {Guilbert}}, \bibinfo {author} {\bibfnamefont {M.}~\bibnamefont {Fontana}},
  \bibinfo {author} {\bibfnamefont {M.}~\bibnamefont {Abarkan}},\ and\ \bibinfo
  {author} {\bibfnamefont {S.}~\bibnamefont {Gille}},\ }\bibfield  {title}
  {\bibinfo {title} {Analysis and control of the dc drift in linbo $ \_
  $\{$3$\}$ $-based mach--zehnder modulators},\ }\href
  {https://ieeexplore.ieee.org/document/5741818} {\bibfield  {journal}
  {\bibinfo  {journal} {Journal of lightwave technology}\ }\textbf {\bibinfo
  {volume} {29}},\ \bibinfo {pages} {1522} (\bibinfo {year}
  {2011})}\BibitemShut {NoStop}%
\bibitem [{\citenamefont {Lihachev}\ \emph {et~al.}(2023)\citenamefont
  {Lihachev}, \citenamefont {Bancora}, \citenamefont {Snigirev}, \citenamefont
  {Tian}, \citenamefont {Riemensberger}, \citenamefont {Shadymov},
  \citenamefont {Siddharth}, \citenamefont {Attanasio}, \citenamefont {Wang},
  \citenamefont {Visani} \emph {et~al.}}]{lihachev2023frequency}%
  \BibitemOpen
  \bibfield  {author} {\bibinfo {author} {\bibfnamefont {G.}~\bibnamefont
  {Lihachev}}, \bibinfo {author} {\bibfnamefont {A.}~\bibnamefont {Bancora}},
  \bibinfo {author} {\bibfnamefont {V.}~\bibnamefont {Snigirev}}, \bibinfo
  {author} {\bibfnamefont {H.}~\bibnamefont {Tian}}, \bibinfo {author}
  {\bibfnamefont {J.}~\bibnamefont {Riemensberger}}, \bibinfo {author}
  {\bibfnamefont {V.}~\bibnamefont {Shadymov}}, \bibinfo {author}
  {\bibfnamefont {A.}~\bibnamefont {Siddharth}}, \bibinfo {author}
  {\bibfnamefont {A.}~\bibnamefont {Attanasio}}, \bibinfo {author}
  {\bibfnamefont {R.~N.}\ \bibnamefont {Wang}}, \bibinfo {author}
  {\bibfnamefont {D.}~\bibnamefont {Visani}}, \emph {et~al.},\ }\bibfield
  {title} {\bibinfo {title} {Frequency agile photonic integrated external
  cavity laser},\ }\href {https://arxiv.org/abs/2303.00425} {\bibfield
  {journal} {\bibinfo  {journal} {arXiv preprint arXiv:2303.00425}\ } (\bibinfo
  {year} {2023})}\BibitemShut {NoStop}%
\bibitem [{\citenamefont {Tian}\ \emph {et~al.}(2020)\citenamefont {Tian},
  \citenamefont {Liu}, \citenamefont {Dong}, \citenamefont {Skehan},
  \citenamefont {Zervas}, \citenamefont {Kippenberg},\ and\ \citenamefont
  {Bhave}}]{tian2020hybrid}%
  \BibitemOpen
  \bibfield  {author} {\bibinfo {author} {\bibfnamefont {H.}~\bibnamefont
  {Tian}}, \bibinfo {author} {\bibfnamefont {J.}~\bibnamefont {Liu}}, \bibinfo
  {author} {\bibfnamefont {B.}~\bibnamefont {Dong}}, \bibinfo {author}
  {\bibfnamefont {J.~C.}\ \bibnamefont {Skehan}}, \bibinfo {author}
  {\bibfnamefont {M.}~\bibnamefont {Zervas}}, \bibinfo {author} {\bibfnamefont
  {T.~J.}\ \bibnamefont {Kippenberg}},\ and\ \bibinfo {author} {\bibfnamefont
  {S.~A.}\ \bibnamefont {Bhave}},\ }\bibfield  {title} {\bibinfo {title}
  {Hybrid integrated photonics using bulk acoustic resonators},\ }\href
  {https://doi.org/10.1038/s41467-020-16812-6} {\bibfield  {journal} {\bibinfo
  {journal} {Nature Communications}\ }\textbf {\bibinfo {volume} {11}},\
  \bibinfo {pages} {3073} (\bibinfo {year} {2020})}\BibitemShut {NoStop}%
\bibitem [{\citenamefont {Baveja}\ \emph {et~al.}(2010)\citenamefont {Baveja},
  \citenamefont {Maywar}, \citenamefont {Kaplan},\ and\ \citenamefont
  {Agrawal}}]{baveja2010self}%
  \BibitemOpen
  \bibfield  {author} {\bibinfo {author} {\bibfnamefont {P.~P.}\ \bibnamefont
  {Baveja}}, \bibinfo {author} {\bibfnamefont {D.~N.}\ \bibnamefont {Maywar}},
  \bibinfo {author} {\bibfnamefont {A.~M.}\ \bibnamefont {Kaplan}},\ and\
  \bibinfo {author} {\bibfnamefont {G.~P.}\ \bibnamefont {Agrawal}},\
  }\bibfield  {title} {\bibinfo {title} {Self-phase modulation in semiconductor
  optical amplifiers: impact of amplified spontaneous emission},\ }\href
  {https://ieeexplore.ieee.org/document/5518533} {\bibfield  {journal}
  {\bibinfo  {journal} {IEEE journal of quantum electronics}\ }\textbf
  {\bibinfo {volume} {46}},\ \bibinfo {pages} {1396} (\bibinfo {year}
  {2010})}\BibitemShut {NoStop}%
\bibitem [{\citenamefont {Zhang}\ \emph {et~al.}(2023)\citenamefont {Zhang},
  \citenamefont {Li}, \citenamefont {Riemensberger}, \citenamefont {Lihachev},
  \citenamefont {Huang},\ and\ \citenamefont
  {Kippenberg}}]{zhang2023fundamental}%
  \BibitemOpen
  \bibfield  {author} {\bibinfo {author} {\bibfnamefont {J.}~\bibnamefont
  {Zhang}}, \bibinfo {author} {\bibfnamefont {Z.}~\bibnamefont {Li}}, \bibinfo
  {author} {\bibfnamefont {J.}~\bibnamefont {Riemensberger}}, \bibinfo {author}
  {\bibfnamefont {G.}~\bibnamefont {Lihachev}}, \bibinfo {author}
  {\bibfnamefont {G.}~\bibnamefont {Huang}},\ and\ \bibinfo {author}
  {\bibfnamefont {T.~J.}\ \bibnamefont {Kippenberg}},\ }\bibfield  {title}
  {\bibinfo {title} {Fundamental charge noise in electro-optic photonic
  integrated circuits},\ }\href {https://arxiv.org/abs/2308.15404} {\bibfield
  {journal} {\bibinfo  {journal} {arXiv preprint arXiv:2308.15404}\ } (\bibinfo
  {year} {2023})}\BibitemShut {NoStop}%
\bibitem [{\citenamefont {Ahn}\ and\ \citenamefont
  {Kim}(2007)}]{ahn2007analysis}%
  \BibitemOpen
  \bibfield  {author} {\bibinfo {author} {\bibfnamefont {T.-J.}\ \bibnamefont
  {Ahn}}\ and\ \bibinfo {author} {\bibfnamefont {D.~Y.}\ \bibnamefont {Kim}},\
  }\bibfield  {title} {\bibinfo {title} {Analysis of nonlinear frequency sweep
  in high-speed tunable laser sources using a self-homodyne measurement and
  hilbert transformation},\ }\href
  {opg.optica.org/ao/fulltext.cfm?uri=ao-46-13-2394&id=131884} {\bibfield
  {journal} {\bibinfo  {journal} {Applied optics}\ }\textbf {\bibinfo {volume}
  {46}},\ \bibinfo {pages} {2394} (\bibinfo {year} {2007})}\BibitemShut
  {NoStop}%
\bibitem [{\citenamefont {Rothman}(2021)}]{rothman2021history}%
  \BibitemOpen
  \bibfield  {author} {\bibinfo {author} {\bibfnamefont {L.~S.}\ \bibnamefont
  {Rothman}},\ }\bibfield  {title} {\bibinfo {title} {History of the hitran
  database},\ }\href {https://www.nature.com/articles/s42254-021-00309-2}
  {\bibfield  {journal} {\bibinfo  {journal} {Nature Reviews Physics}\ }\textbf
  {\bibinfo {volume} {3}},\ \bibinfo {pages} {302} (\bibinfo {year}
  {2021})}\BibitemShut {NoStop}%
\bibitem [{\citenamefont {Welch}(1967)}]{welch1967use}%
  \BibitemOpen
  \bibfield  {author} {\bibinfo {author} {\bibfnamefont {P.}~\bibnamefont
  {Welch}},\ }\bibfield  {title} {\bibinfo {title} {The use of fast fourier
  transform for the estimation of power spectra: A method based on time
  averaging over short, modified periodograms},\ }\href
  {https://doi.org/10.1109/TAU.1967.1161901} {\bibfield  {journal} {\bibinfo
  {journal} {IEEE Transactions on Audio and Electroacoustics}\ }\textbf
  {\bibinfo {volume} {15}},\ \bibinfo {pages} {70} (\bibinfo {year}
  {1967})}\BibitemShut {NoStop}%
\bibitem [{\citenamefont {Domenico}\ \emph {et~al.}(2010)\citenamefont
  {Domenico}, \citenamefont {Schilt},\ and\ \citenamefont
  {Thomann}}]{di2010simple}%
  \BibitemOpen
  \bibfield  {author} {\bibinfo {author} {\bibfnamefont {G.~D.}\ \bibnamefont
  {Domenico}}, \bibinfo {author} {\bibfnamefont {S.}~\bibnamefont {Schilt}},\
  and\ \bibinfo {author} {\bibfnamefont {P.}~\bibnamefont {Thomann}},\
  }\bibfield  {title} {\bibinfo {title} {Simple approach to the relation
  between laser frequency noise and laser line shape},\ }\href
  {https://doi.org/10.1364/AO.49.004801} {\bibfield  {journal} {\bibinfo
  {journal} {Appl. Opt.}\ }\textbf {\bibinfo {volume} {49}},\ \bibinfo {pages}
  {4801} (\bibinfo {year} {2010})}\BibitemShut {NoStop}%
\end{thebibliography}%
	
\end{document}